%% file: main.tex
\documentclass[10pt,journal,cspaper,compsoc]{IEEEtran}
\ifCLASSOPTIONcompsoc
\else
\fi
\ifCLASSINFOpdf
\else
\fi
\usepackage{graphicx}
\usepackage{subfigure}

\usepackage{amsmath}
\usepackage{amssymb}
\usepackage{color}
\usepackage{url}
\usepackage{graphicx}
\usepackage{dcolumn}
\usepackage{bm}
\usepackage{hyperref}
\usepackage{multirow}

\usepackage[english]{babel}
\makeatletter
\adddialect\l@en\l@english
\makeatother

\usepackage{xcolor}

\begin{document}
%
\title{Spreading processes in Multilayer Networks}

\author{
\null
\authorblockN{Mostafa Salehi}\\
\authorblockA{University of Bologna, Italy and University of Tehran, Iran\\
Corresponding author Email: mostafa\_salehi@ut.ac.ir}\\
\null
\and
\authorblockN{Rajesh Sharma}\\
\authorblockA{University of Bologna, Italy\\
Email: rajesh.sharma@unibo.it}\\
\null
\and
\authorblockN{Moreno Marzolla}\\
\authorblockA{University of Bologna, Italy\\
Email: moreno.marzolla@unibo.it}\\
\null
\and
\authorblockN{Matteo Magnani}\\
\authorblockA{Uppsala University, Sweden\\
Email: matteo.magnani@it.uu.se}\\
\null
\and
\authorblockN{Payam Siyari}\\
\authorblockA{Sharif University of Technology, Iran\\
Email: siyari@ce.sharif.edu}\\
\null
\and
\authorblockN{Danilo Montesi}\\
\authorblockA{University of Bologna, Italy\\
Email: danilo.montesi@unibo.it}\\
\thanks{This work has been partly funded by FIRB project Information monitoring, propagation analysis and community detection in Social Network Sites.}
}

\IEEEcompsoctitleabstractindextext{%

\begin{abstract}
Several systems can be modeled as sets of interconnected networks or networks with multiple types of connections, here generally called multilayer networks. Spreading processes such as information propagation among users of an online social networks, or the diffusion of pathogens among individuals through their contact network, are fundamental phenomena occurring in these networks. However, while information diffusion in single networks has received considerable attention from various disciplines for over a decade, spreading processes in multilayer networks is still a young research area presenting many challenging research issues. In this paper we review the main models, results and applications of multilayer spreading processes and discuss some promising research directions.

\end{abstract}



\begin{keywords}
Multilayer Network, Multiplex, Interconnected, Spreading processes, Diffusion
\end{keywords}}

\maketitle



\IEEEdisplaynotcompsoctitleabstractindextext


%
\IEEEpeerreviewmaketitle

\input{introduction}

\input{framework}

\input{modelling}

\input{observing}
\input{application}

\input{conclusion}


%


\end{document}

%% file: introduction.tex
\section{Introduction}\label{sec:intro}
Many real-world systems can be modeled as networks, i.e., sets of interconnected entities. In some cases the connections between these entities represent communication channels: they indicate that information items present at one of the entities can be transferred, or propagated, to some neighbor entities. A typical example is represented by online social networks, where information can move from one user account to the other through e.g.\@ \emph{friendship} or \emph{following} connections, but several other scenarios exist where the nodes of the network are not human beings (e.g., computer networks and the so-called Internet of things)
and the items traversing the network are not text messages but for instance viral agents, rumors, behaviors, pathogens or digital viruses. These are all examples of spreading processes.

Studying the diffusion of pathogens has a long history in biological systems, and a robust analytic framework has developed in epidemiology for modeling this type of spreading processes~\cite{Bailley1975, Anderson1992}. With the advent of network science, the traditional epidemic models were extended to incorporate the structure of the underlying network~\cite{Pastor-Satorras2001} and utilized to study network epidemics~\cite{Moreno2002, Newman2002, Chakrabarti2008, mieghem2009}. Such modeling has recently attracted considerable attention in spreading processes over communication systems~\cite{Kleinberg2007, Wang2009, Meisel2010} and online social communities~\cite{GuilleSurveyInfoDiss, Borge-Holthoefer2013a}.

However, although spreading processes on networks have been thoroughly studied during over the last decade~\cite{Pastor-Satorras2014}, real spreading phenomena are seldom constrained into a single network (called monoplex network). This is evident in online information propagation, where the process of switching network while sharing information on social media has become a basic functionality explicitly provided by many platforms. Another example is represented by the diffusion of epidemics propagated by human beings traveling via multiple transport networks (airplanes, trains, etc.).

\begin{figure}[!tbp]
  \caption{Three main dimensions for analyzing spreading processes in multilayer networks: (i) how to model the spreading processes, (ii) what results we can obtain using these models and (iii) how these results can be exploited in real applications}
  \centering
    \includegraphics[width=0.4\textwidth]{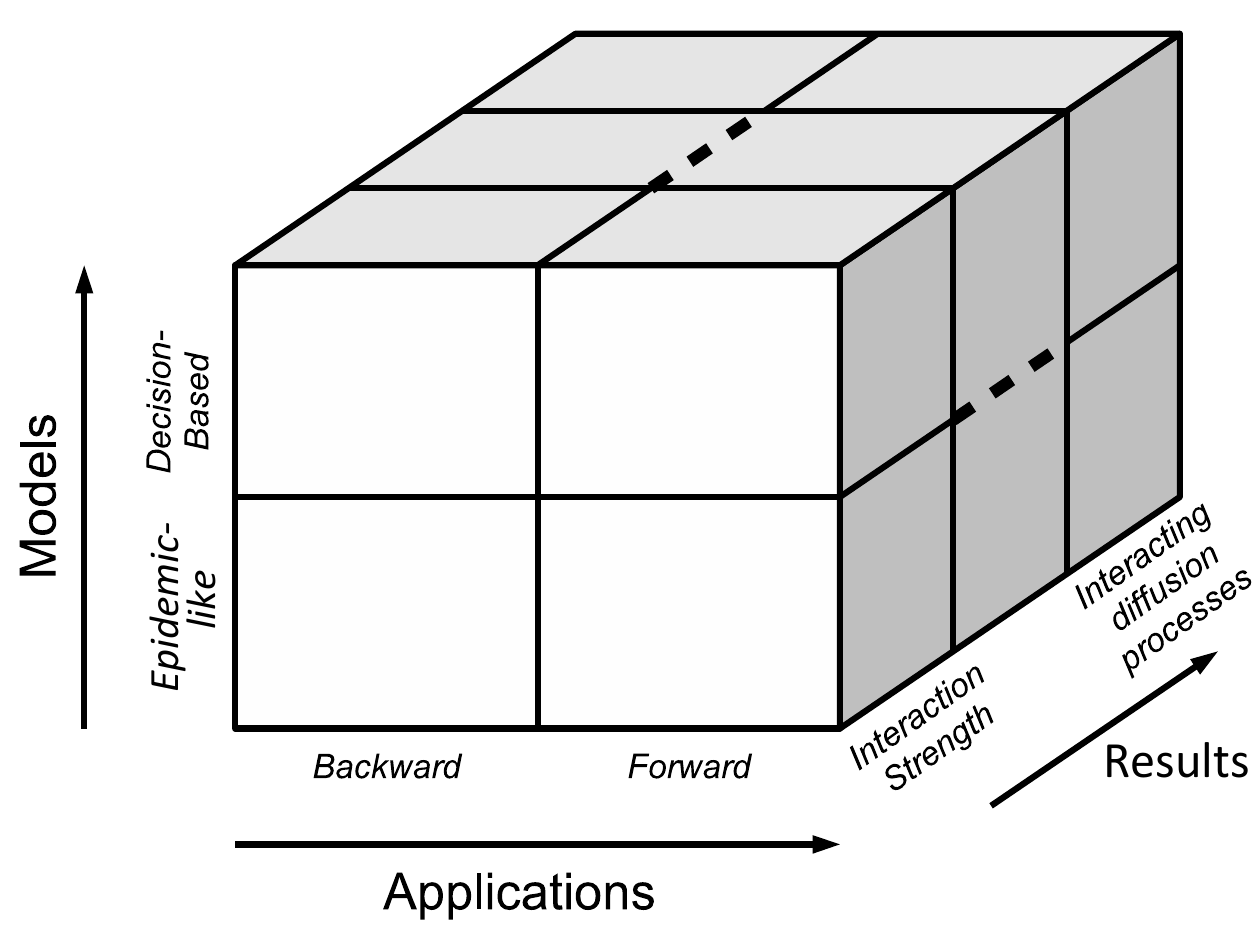}\label{fig:3Dimen}
\end{figure}

In this paper we focus on the practically relevant topic of \emph{spreading processes} in \emph{multilayer networks}
a generic term that we use to refer to a number of models involving multiple networks, called \emph{interconnected networks} \cite{PhysRevE.88.022801}, or multiple types of relationships, called \emph{multiplex networks} \cite{Lee2014}.
Multilayer networks are also known as interdependent \cite{Buldyrev2010, Danziger2014, Kenett2014}, multidimensional \cite{Berlingerio2012}, multiple \cite{Magnani2013}, multisliced \cite{Mucha2010}, multilevel \cite{Wasserman1994,Wang2013a} networks, and networks of networks \cite{D'Agostino2014, Gao2014}.
Historically, networks were first inspected from the multilayer perspective by sociologists in works such as \cite{Roethlisberger1939} in the late 1930s, and research continued in subsequent years \cite{Gluckman1955, Mitchell1969, Verbrugge1979, Breiger1986, Pattison1999}. Multilayer networks have attracted interest again in recent years~\cite{Kivela2013, Boccaletti2014,D'Agostino2014}. 
For a general overview of multilayer networks, the reader is referred to review articles~\cite{Kivela2013, Boccaletti2014} and the book~\cite{D'Agostino2014}.

In addition to spreading processes (which is the focus of this paper), numerous other types of diffusion processes on multilayer networks have been studied, including cascading failures \cite{Buldyrev2010,  Shao2011, Brummitt2012, Shin2014}, cooperative behavior  \cite{gomezPRE2012, WangSP13, Jiang2013, Santos2014}, and synchronization \cite{Barreto2008, Bogojeska2013, Aguirre2014}.

When only single networks are involved, it is well known that for all the processes above the structure of the network plays an important role on the outcomes of the process. For example, behavior spreading can stall when it enters a tightly-knit community within the network \cite{Easley2010}. The same is true when multilayer networks are involved, but the effect of the layer structures and their interdependence may differ from the single-network case. Today, the study of spreading processes in multilayer networks is a young and rapidly evolving research area facing challenging issues. In this paper we provide a homogeneous overview of current results on the effect of multiple layers and other network features on the diffusion of different types of items, and identify unexplored areas.

To this end, we analyze the topic of spreading processes in multilayer networks according to three main aspects: (i) how spreading processes can be modeled (Section \ref{sec:modeling}), (ii) what results can be obtained from these models (Section \ref{sec:obsr}) and (iii) how these results can be exploited in real applications (Section \ref{sec:app}). These aspects are summarized in Figure \ref{fig:3Dimen}. The paper follows the same structure: after introducing the basic concepts (multilayer networks, spreading processes in multilayer networks and variables used to study these phenomena) we devote one section to each of the aforementioned aspects. Finally we present a set of open problems in the area that in our opinion still require significant research efforts (Section \ref{sec:con}).

%% file: framework.tex
\section{Preliminaries}\label{sec:framework}

\begin{table*} [t]
\centering
\caption{Notation}
\begin{center}
\begin{tabular}{| l | p{10cm} |}
\hline
$V$ & The set of nodes in a multilayer network\\
$L$ & The set of layers in a multilayer network\\
$n$ & The number of nodes in a multilayer network\\
$(u,l_u)$ & Corresponds to node $u$ on layer $l_u$ in a multilayer network\\
$((u,l_u),(v,l_v))$ & The tuple representing an edge between node $u$ on layer $l_u$ and node $v$ on layer $l_v$ in a multilayer network\\
$C$ & An information cascade\\
$(u,l_u,v,l_v,t)_C$ & The entries of the set denoted by the information cascade $C$\\
$D$ & A (multilayer) diffusion network\\ \hline
\end{tabular}
\end{center}
\label{Tbl:NotationTable}
\end{table*}

In this section we introduce the concepts of multilayer network and spreading processes in multilayer networks, and the main methods and variables used to study these processes.

\begin{figure*}[t]
\centering
{
\subfigure[
\emph{A Multilayer Network}
]{\includegraphics[trim = 40mm 50mm 40mm 1mm, scale=0.4]{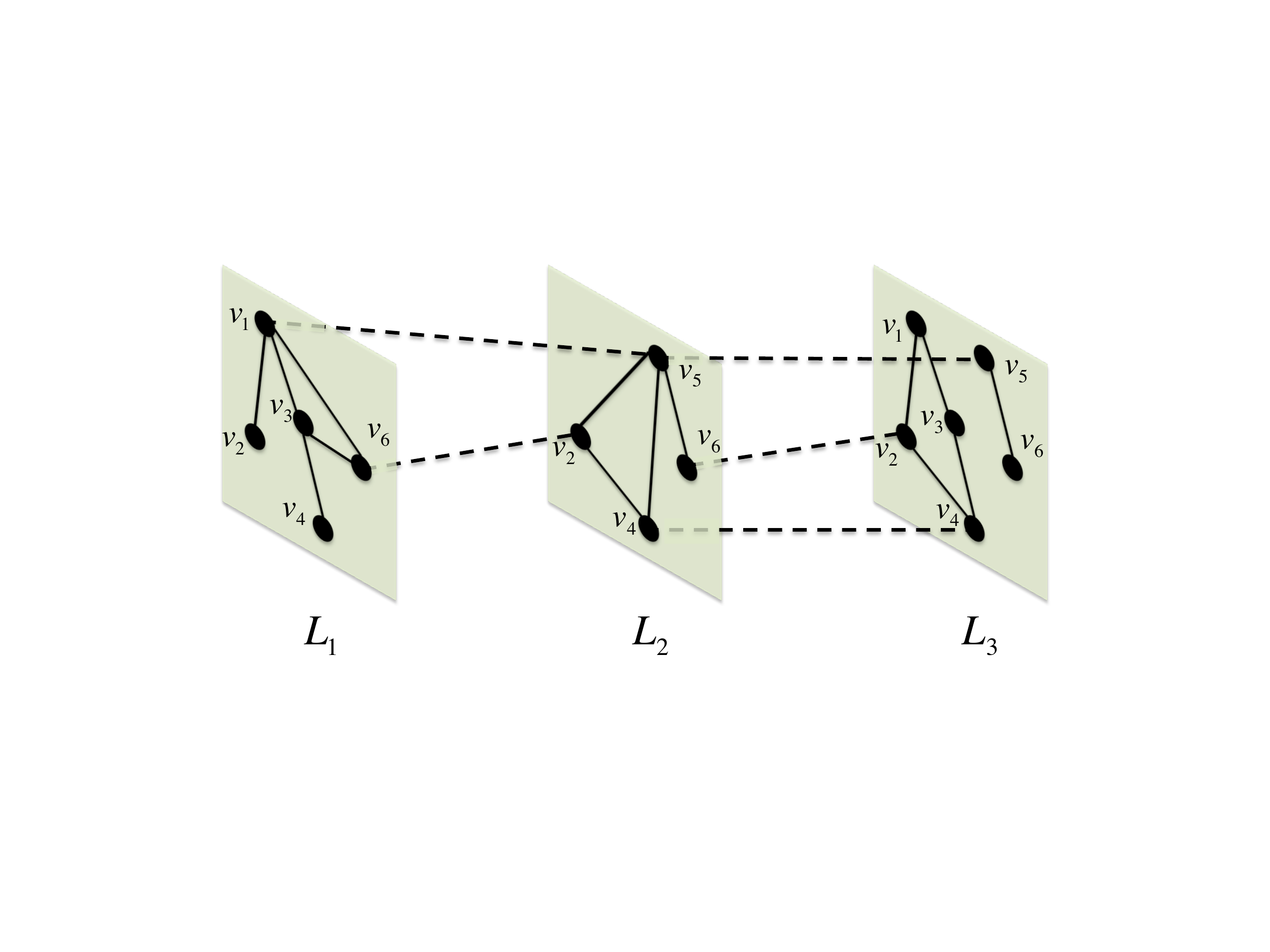}
\label{fig:fig1a}}
\subfigure[
\emph{A spreading process starting from seed node $(v_4,l_2)$ on the multilayer network in Figure~\ref{fig:fig1a}. This is also called a \emph{cascade}, and we refer to it as $C_1$}
]{\includegraphics[trim = 10mm 50mm 40mm 1mm, scale=0.4]{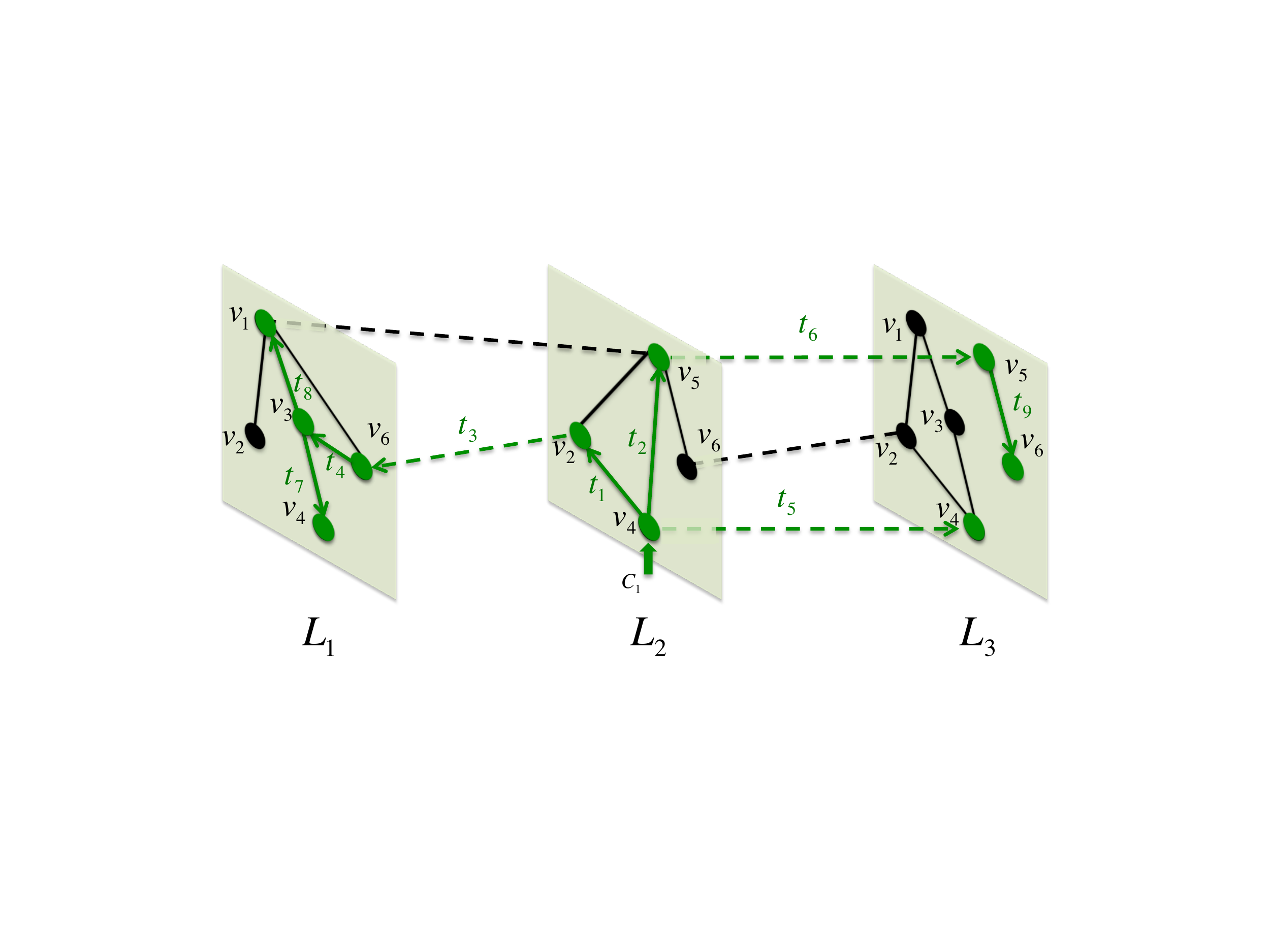}
\label{fig:fig1b}}
\subfigure[
\emph{A cascade $C_2$ generated at seed node $(v_4,l_1)$}
]{\includegraphics[trim = 40mm 50mm 40mm 1mm, scale=0.4]{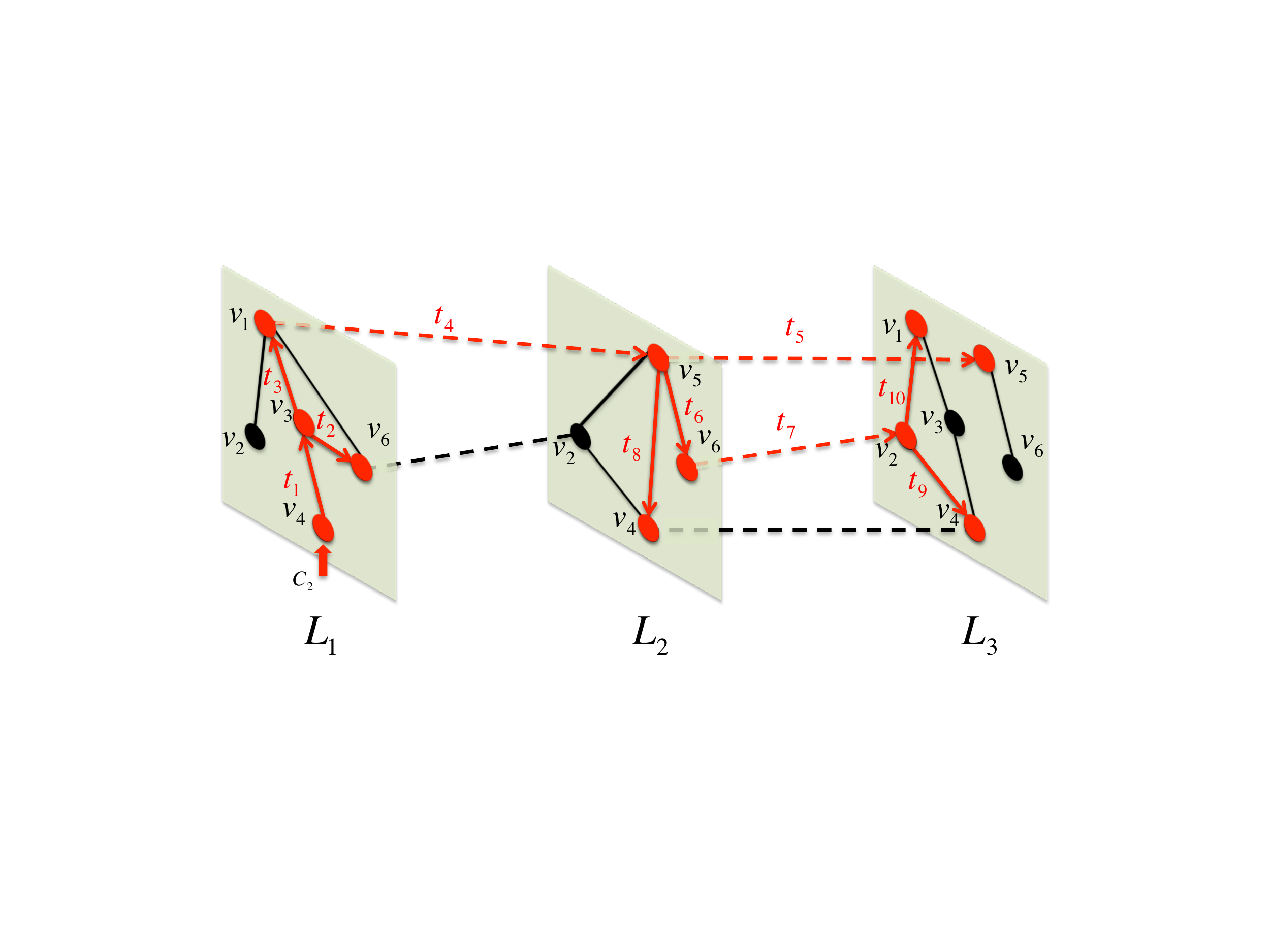}
\label{fig:fig1c}}
\subfigure[
\emph{The diffusion network resulting by the aggregation of cascades $C_1$ and $C_2$}
]{\includegraphics[trim = 10mm 50mm 40mm 1mm, scale=0.4]{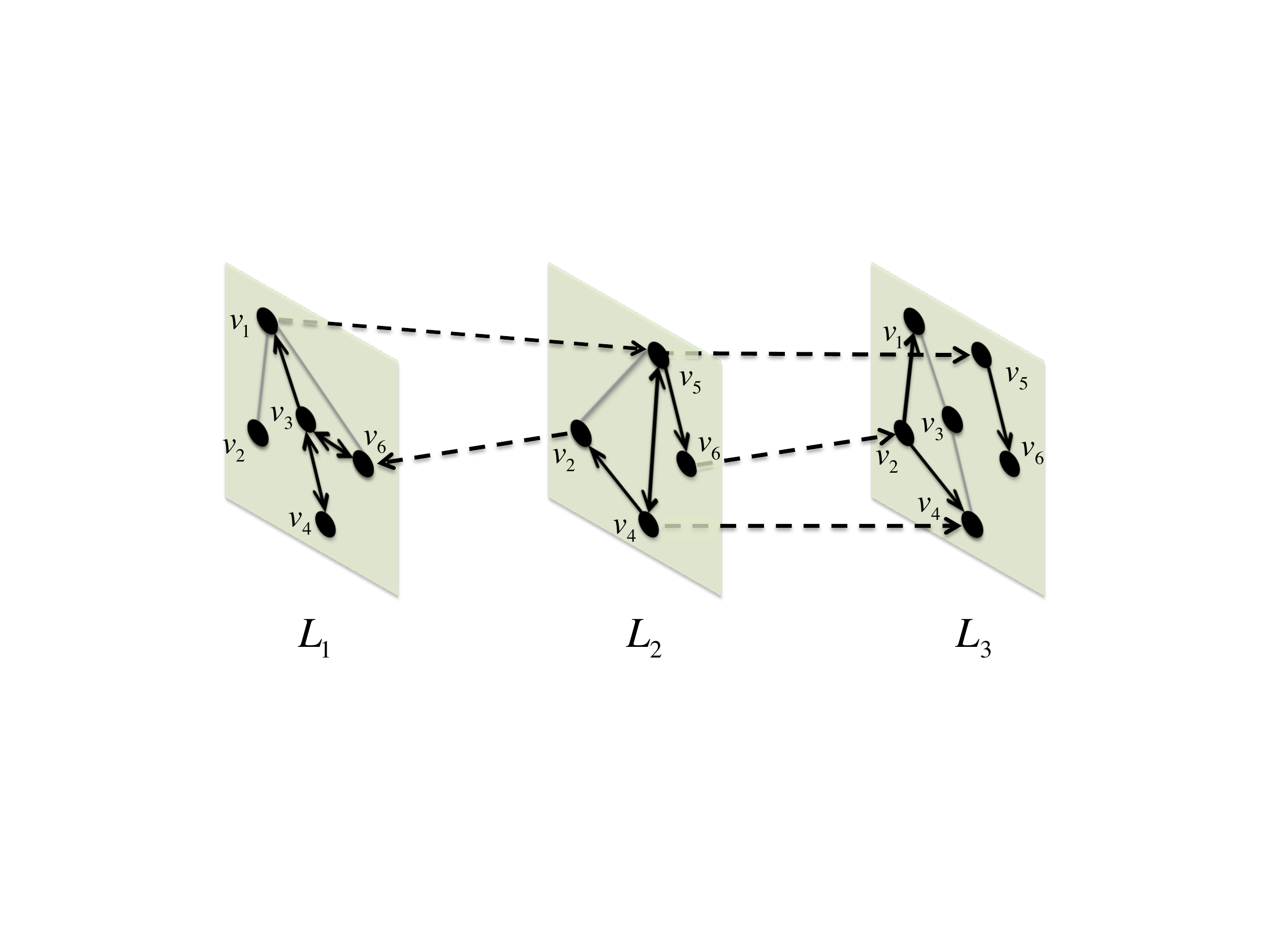}
\label{fig:fig1d}}
}
\caption{Spreading processes in multilayer networks. {\bf(a)} The underlying multilayer network. In this example edges are undirected and solid and dashed lines represent intra-layer and inter-layer edges, respectively. Notice that nodes have the same name across layers and may not be present in some of them. {\bf(b,c)} Example cascades $C_1$ and $C_2$ diffusing over the underlying multilayer network starting from nodes $(v_4,l_2)$ and $(v_4, l_1)$, respectively. The arrows represent the direction of spreading processes: dashed arrows represent inter-layer information propagation, edges $\left((v_5,l_2),(v_5,l_3)\right)$ and $\left((v_4,l_2),(v_4,l_3)\right)$ in $C_1$ and edge $\left((v_5,l_2),(v_5,l_3)\right)$ in $C_2$ are examples of \emph{same-node inter-layer} diffusion, edge $\left((v_2,l_2),(v_6,l_1)\right)$ in $C_1$ and edges $\left((v_1,l_1),(v_5,l_2)\right)$ and $\left((v_6,l_2),(v_2,l_3)\right)$ in $C_2$ are examples of \emph{other-node inter-layer} diffusion. The other edges in $C_1$ and $C_2$ are examples of  \emph{other-node intra-layer} diffusion. Nodes do not need to participates in \emph{same-node inter-layer} diffusion. For example, node $v_4$ in $L_2$, spreads the information to Layer $L_3$ but not to $L_1$. Also, node $v_4$ in $L_1$ does not spread to layers $L_2$ and $L_3$. {\bf(d)} The subgraph resulting from the aggregation of cascades $C_1$ and $C_2$ forms the diffusion network.}
\label{fig:fig1}
\end{figure*}

We assume that the reader is already familiar with the concept of
\emph{graph}: a graph $G = (V, E)$ is a finite set of nodes (vertices)
$V$ and a set of (ordered or unordered) pairs $E \subseteq V \times
V$. A monoplex network is a (usually directed) graph.
A multilayer network is a data structure made of multiple layers, where each layer is a monoplex network.
Here we use the general mathematical framework defined in~\cite{Kivela2013} (see also~\cite{DeDomenico2013} as the first attempts to provide multilayer network science with a consistent mathematical representation).
In this framework, the same nodes can appear in multiple layers and nodes on different layers can be connected to each other.
As an example, in Figure~\ref{fig:fig1a} the pairs $(v_4,l_2)$, $(v_4,l_3)$ and $(v_5,l_2)$ identify specific nodes in the different layers, in particular node $v_4$ on layers $l_2$ and $l_3$ and node $v_5$ on layer $l_2$. Layer $l_2$ corresponds to a monoplex network, with simple edges like ($(v_4,l_2)$, $(v_5,l_2)$) -- or just $(v_4,v_5$) if we know we are referring to layer $l_2$. In addition, we can have edges between layers, e.g., $((v_4,l_2)$,  $(v_4,l_3))$. In the context of this paper, edges model e.g.~communication channels: in Figure~\ref{fig:fig1a} if $v_6$ has some information on layer $l_2$ s/he can propagate it to $v_5$ on the same layer or send it to $v_2$ on layer $l_3$.

Building on this basic model several attributes can be added to nodes and edges. For example, we can introduce a temporal dimension and make a distinction between node $v_4$ on layer $l_2$ at time $t_0$ $(v_4,l_2,t_0)$ and the same node at time $t_1$ $(v_4,l_2,t_1)$, and we can then add edges among these extended nodes, like $((v_4,l_2,t_0), (v_4,l_2,t_1))$, or $((v_6,l_2,t_0), (v_2,l_3,t_0))$.
In~\cite{Kivela2013} these attributes (layer, time, etc.) are called \emph{aspects}. The notation introduced in the remainder of the paper is summarized in Table~\ref{Tbl:NotationTable}.

Generally, we can consider two extreme cases for the nodes in a multilayer network. In one case all layers contain the same set of nodes,
as in the case of individuals that may take part to different online social networks (i.e., layers) at the same time; in this example, all layers consist of approximately the same nodes, with the only exception of nodes representing those users that do not have accounts on some specific social networks.
A multilayer network where all layers contain almost the same set of nodes is called a \emph{multiplex network} in the literature~\cite{Lee2014}.
At the other extreme, each node of a multilayer network may belong to exactly one layer,
resulting in a data structure sometimes called \emph{interconnected} \cite{PhysRevE.88.022801} (or
\emph{interdependent} \cite{Buldyrev2010, Danziger2014, Kenett2014})
network; in interconnected networks self-interactions across different layers are therefore not possible. In a different perspective, interconnected networks can be viewed as ``interconnected communities within a single, larger network'' \cite{Dickison2012a}.
As an example of interconnected networks we may consider the power and communication infrastructures, where the functionality of each one of the two networks depends on the other, and failure of particular nodes in either of the networks compromises the operation of the other network~\cite{Buldyrev2010}.

As said, connections between nodes on the same or different layers represent channels through which different types of items can propagate, giving rise to spreading processes.
In general, spreading process can refer to the diffusion of pathogens, rumors, behaviors, or the coverage of a news-headline in different newsgroups and weblogs. Although all the above contexts share some common aspects, there are specific features differentiating the various types of spreading processes. For example, in the case of spreading of some behavior in a community, people usually choose which behavior to adopt. On the other hand, in the case of epidemics there is no decision made by the individuals who are infected.
These topics are thoroughly covered in Section~\ref{sec:modeling}. In the current section, we  present the key concepts that may arise in the analysis of spreading processes.

The evidence left from the diffusion of a particular piece of
information over a monoplex network is called
\emph{(information) cascade}~\cite{Eslami2011,Banos2013b}.  This
concept can be extended for multilayer networks~\cite{Li2014}, as
shown in Figures~\ref{fig:fig1b} and~\ref{fig:fig1c}. It also
generates an implicit network as shown in Figure~\ref{fig:fig1d}. Therefore we will sometimes distinguish between a
\textbf{diffusion network} (i.e., the actual connections traversed
during the diffusion process) and an \textbf{underlying (multilayer) network}. A diffusion network is defined by the sequence of nodes
traversed by a certain piece of information or other item. In a
multilayer network a cascade can be represented as a set of tuples
$(u, l_u, v, l_v, t)$ where $t$ represents the timestamp when the
propagated item passed from node $u$ in layer $l_u$ to node $v$ in
layer $l_v$. We call \emph{seed} the first node of the tuple with the
minimum timestamp. While this is the minimum amount of information
needed to meaningfully describe a spreading process, specific models
reviewed in Section~\ref{sec:modeling} augment these tuples with
additional parameters (i.e., a state space and a set of rules for
state-transition) providing more details about the cascade.

\begin{figure}[tbp]
\centering\includegraphics{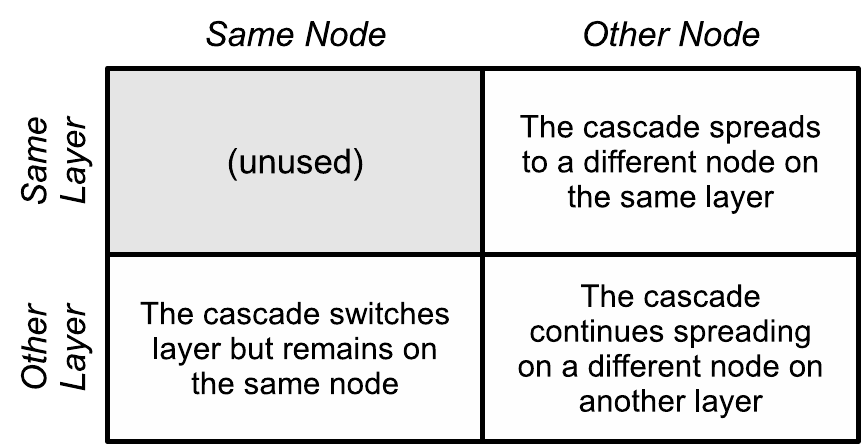}
\caption{Different possibilities for spreading an item from one layer to another in a multilayer network}\label{fig:traversing}
\end{figure}

One of the important ideas in the context of spreading processes in
multilayer networks is the fact that items can also spread from one layer
to another. In general there are four possibilities for an item to
traverse a multilayer network (see Figure~\ref{fig:traversing}):
\textbf{same-node inter-layer}, when the cascade switches layer but
remains on the same node, e.g., when a Facebook post is shared on
Twitter by the author of the same post; \textbf{other-node
  inter-layer}, when a cascade continues spreading to another node in
another layer, e.g., exchanging mails between users with different mail accounts (e.g., gmail and yahoo).
In third type, \textbf{other-node intra-layer}, the cascade continues
spreading through the same layer, e.g., retweeting a post in Twitter.
It is worth noting that inter-layer diffusion may involve a
\emph{layer-crossing overhead} (which is also called
\emph{layer-switching overhead}) \cite{Min2013}. The fourth
combination, \textbf{(iv) same-node intra-layer}, is generally not
considered meaningful and therefore omitted in all the diffusion
studies we have considered. \cite{DBLP:conf/asonam/MagnaniR11}
introduced a model where the same individuals can have multiple nodes
(e.g., accounts) on the same network. In this case, information might
flow from one individual to the same individual, from one account to
the other. However, to the best of our knowledge this model has not
been used to study diffusion processes yet.

Figure~\ref{fig:fig1} presents a summary of the concepts of underlying
multilayer network, different types of information cascades and the
resulting diffusion network. The corresponding terminology introduced
so far is indicated in Table~\ref{Tbl:TermsTable} for quick reference.

\begin{table*} [tbp]
	\caption{Basic terminology}
	\begin{center}
	\begin{tabular}{| l | p{13cm} |}
	\hline
		Term & Explanation \\ \hline
		Multilayer Network & General term for a network with multiple layers \\
		Multiplex Network & Multilayer network with approximately the same set of nodes across all layers \\ 
		Interconnected Network & Multilayer network in which the nodes are of different types \\ 
		Monoplex Network & Network with a single layer \\
		Information Cascade & The trace left by the spread of information \\
		Seed(s) & The node(s) from which an information cascade starts spreading \\ 
		Diffusion Network & The subgraph resulted from the aggregation of covered subgraphs of information cascades \\
		Intra/Inter-Layer Diffusion & The spread of an information cascade within/between layers of a multilayer network \\ 
		 \hline
\end{tabular}
\end{center}
\label{Tbl:TermsTable}
\end{table*}

Studies based on spreading processes can be categorized into three
types.  \textbf{Empirical studies} involve the analysis of real
datasets, either complete or sampled~\cite{Mehdiabadi2012,
  Mehdiabadi2012a}. These studies would be extremely important to
understand the real dynamics of information diffusion. However, to the
best of our knowledge there are so far no works based on real datasets
of information diffusion in multilayer networks.  Unlike cases
involving a monoplex network
\cite{Leskovec07blogs,Magnani2010a,WangWWW2011,BakshyWWW2012,magnani2013factors},
it is non-trivial to analyze the process in multilayer networks. In
\cite{Gjoka2011,DBLP:conf/asonam/MagnaniR11} the authors use sampling
methods for collecting data from multiple online social networks, but
do not have access to information cascades.  Given the difficulty in
collecting real datasets including both the diffusion process and the
underlying network where diffusion takes place, the totality of
existing works on multilayer diffusion are either
\textbf{simulation-based studies}, where a synthetic~\cite{epstein08,
  Min2013, Cozzo2013a, Buono2013, Li2014, Qian2012, Zhao2014770} or
real
\cite{VidaGC13,rsharmaDecInfoDissMulDim,IdnCritNodMulLayNwUnMulVecMalAtk,
  Wei2012, CompMemProOnNwPersp, Budak2011, Cozzo2013a, Wang2011a}
network is used to host artificial diffusion processes, and
\textbf{analytic studies} working with mathematical models of
information diffusion~\cite{Yagan2013, Buono2013,
  Saumell-Mendiola2012, Cozzo2013a, Buono2013, Sanz2014}.

Both simulation and analytic studies are based on the observation of
the behavior of specific variables of interest depending on some input
parameters. A basic parameter included in all studies is the
\textbf{transmissibility probability}, indicating the probability of
transmitting an item (i.e., transferring an infection, passing a
message, spreading some rumor) from one node to the other
\cite{Min2013}. Care should be taken in providing a unique definition
of this concept, because variations can be found in the
literature. For example, in~\cite{Yagan2013} transmissibility is
considered to be the mean value of this probability computed among all
neighbor nodes, and~\cite{Yagan2013,Min2013} distinguish between
different kinds of transmissibility -- homogeneous and
heterogeneous. Other important input parameters in simulation studies
are the \textbf{type of underlying networks} (e.g., random
\cite{Buono2013}, scale-free~\cite{Yagan2013}, small-world
\cite{Li2014}, etc) and the \textbf{relationships between different
  layers}~\cite{PhysRevE.88.022801, Wang2011a, Saumell-Mendiola2012}
(e.g., the correlation between node degrees
\cite{Saumell-Mendiola2012}).

We conclude this section by presenting the main dependent variables
used in different diffusion studies. The so-called \textbf{epidemic
  threshold} \cite{Bailley1975, Hethcote2000} is one of the key
observations in epidemic-like models (refer to
Section~\ref{sec:modeling}), and indicates a value of transmissibility
above which the diffusion involves the whole (or most of the) network,
e.g., the diffusion network is a giant component of the underlying
network. It is known that in monoplex networks the value of the
epidemic threshold is closely related to the largest eigenvalue of the
network's adjacency matrix~\cite{Wang03,Gleeson2011}.  Furthermore,
recent work suggests that the epidemic threshold in a multiplex
network cannot be larger than the epidemic thresholds of individual
layers~\cite{Min2013}. In the context of interacting spreading
  processes in multilayer networks (refer to~\ref{InterProc}), two
  types of thresholds have recently been introduced, called
  \textbf{survival threshold} and \textbf{absolute-dominance
    threshold}: they measure if a diffusion process will survive and
  whether it can completely remove another competing process
  \cite{DarabiSahneh2014}. Another dependent variable is the
\textbf{infection size}, generally defined as the number or fraction
of nodes in the diffusion network, i.e., those reached by the
diffusion process. Similar terms such as \emph{outbreak
  size}~\cite{Hethcote2000} or \emph{cascade size}~\cite{Li2014}
have also been used in the literature to refer to this quantity. The
\textbf{Infection rate}, representing the average rate of being in
contact over a link, is also a frequently studied dependent variable.

While epidemic threshold and infection size are static measures of a
spreading process, some observational variables also take temporal
aspects into account. For example, in some epidemic models an infected
node may recover from the disease or may die and be removed from the
network. As a consequence, the number of \lq\lq infected\rq\rq nodes
changes with time. The percentage of infected nodes (i.e., infection
size) at a specific time is sometimes called \textbf{epidemic
  dynamics}~\cite{Gomez2013}; the \textbf{cascade velocity} measures
how fast an item (e.g., a message) reaches some relevant nodes or a
given number of nodes in a cascade~\cite{Li2014}.  Finally, we can
study the \textbf{survival probability}, which is the probability that
an infection, started from a single node, is still active at
time $t$~\cite{Dickison2012a}. \textbf{Outbreak probability} indicates
the probability that a seed infection gives rise to an epidemic
outbreak~\cite{Min2013,PhysRevE.79.036113}.

\textbf{Recall} and \textbf{precision} are two widely used
  measures in the field of information retrieval and pattern
  recognition. For spreading phenomena over networks, recall can be
  defined as the ratio of the number of relevant nodes in the
  diffusion network divided by the total number of relevant nodes,
  while precision is the ratio of the number of relevant nodes in the
  diffusion network divided by the total number of nodes in the
  diffusion network~\cite{Datta2011GSG}. In this context, relevance is
  an application-specific measure of ``interest'' of a node in the
  item that it is being spread.

Table~\ref{Tbl:VariablesTable} summarizes the definitions of the
aforementioned variables.

\begin{table*} [!htbp]
  \small
	\caption{The main dependent variables used in different diffusion studies}
	
    \begin{tabular}{|c|l|p{11cm}|}
    \hline
    \multirow{1}{*}{Type} & \multirow{1}{*}{Variable Name} & \multirow{1}{*}{Definition} \\
	
    \hline
     \multirow{7}{*}{Static}
     & Transmissibility & The probability of transmitting an item from one node to another.\\
		& Epidemic Threshold & A value of transmissibility above which the diffusion process involves most of the network.\\
       & Survival Threshold &
       Given two interacting spreading processes, the survival threshold is a critical point for effective infection rate of one process above which this process survives~\cite{DarabiSahneh2014}.\\
	    & Absolute-dominance Threshold & Given two interacting spreading processes, the absolute-dominance threshold is a critical point for effective infection rate of the first process such that not only this process survives but also it removes the competing process~\cite{DarabiSahneh2014}.\\
		& Infection Size & The number or fraction of nodes in the diffusion network.\\
		& Cascade Size & The number of infected nodes in a cascade. \\
		& Infection Rate & The average rate of being in contact over a link. \\
	
  \hline
     \multirow{4}{*}{Temporal}
 & Epidemic Dynamics & The percentage of infected nodes (i.e., infection size) at a specific time.\\
		& Cascade Velocity & How fast an item reaches some relevant nodes or a given number of nodes in a cascade.\\ 		
		& Survival Probability & The probability of an infection started from a single infected node being active at a specific time.\\
				
\hline
     \multirow{4}{*}{Target-based}
 		& Outbreak Probability & The chance that a seed infection gives rise to an epidemic outbreak.\\
		& Recall & The ratio of the number of relevant nodes in the diffusion network divided by the total number of relevant nodes.\\
		& Precision & The ratio of number of relevant nodes in the diffusion network divided by the total number of nodes in the diffusion network.\\
		\hline

	\end{tabular}%
  \label{Tbl:VariablesTable}%
\end{table*}%

%% file: modelling.tex
\section{Modeling Spreading Processes in Multilayer Networks}\label{sec:modeling}

As said, it is non-trivial to obtain real data for analyzing spreading
processes in multilayer networks. Therefore, as an alternative
approach, modeling can be used for understanding and analyzing the
dynamics of spreading processes over the networks.  Here, we discuss
various research works which have attempted to model spreading
processes in multilayer networks. We first review and
  categorize existing spreading models (Section~\ref{sec:mod:class})
  and then describe theoretical approaches for the analysis of these
  models (Section~\ref{sec:mod:anal}).

\subsection{Review and classification of existing spreading models}\label{sec:mod:class}
We categorize existing models in two groups: epidemic-like (Section~\ref{sec:mod:epid}) and decision-based (Section~\ref{sec:mod:dec}).

\subsubsection{Epidemic-like models}\label{sec:mod:epid}
In epidemic-like models, generally used for modeling disease and influence spreading, the probability that a node becomes infected by a diffusion process (e.g., disease spreading) is determined by its neighbors or adjacent nodes~\cite{Pastor-Satorras2001}.
Most of the work on modeling the dynamics of diffusion over multilayer networks has used epidemic models such as SIR~\cite{Min2013, Dickison2012a, Qian2012, Zhao2014770, Marceau2011, Cozzo2013a, Yagan2013, Buono2013}, SIS~\cite{PhysRevE.88.022801, Sanz2014, Saumell-Mendiola2012} and SI1I2R~\cite{Wei2012, CompMemProOnNwPersp}.

The dynamics of epidemic spreading according to the SIR and SIS models
are described as a three- and two-state process, respectively. The
diffusion process starts with an initial infected set of nodes, called
seeds. An infected node diffuses the infection (i.e., information,
disease) to a susceptible neighbor with infection rate $\beta$. The
infected nodes can recover after time $\tau$ from the moment of
infection, as in the susceptible-infected-recovered (SIR) model; or
they can change their state back to susceptible as in the
susceptible-infected-susceptible (SIS) model.  Many extensions have
been applied to SIR and SIS models; interested readers can refer to~\cite{Kermack1927,Newman2010} for more details and various
extensions. As one of the most important extensions, Goldenberg et
al.~\cite{Goldenberg2001} proposed a discrete-time version of the SIR
model called \emph{Independent Cascade Model (ICM)}, where time
proceeds in discrete time steps. In this model, each infected node $u$
at time $t$ can infect each of its neighbors. If the infection
succeeds, then neighbor $v$ will become infected at step $t+1$. ICM is
often used in the literature on influence spreading. In
\cite{Budak2011}, the authors extended this model to analyze the
dynamics of multiple cascades over a multiplex network.

In a monoplex network, the probability of transferring an (information) item from one node to another (i.e., transmissibility) can be computed as
$T=1-e^{\lambda}$ in the continuous case~\cite{Newman2002}, where $\lambda$ is the effective infection rate. Also, $\lambda$ = $\beta \tau$, where $\beta$ is the infection rate and $\tau$ represents the time for which a node remains infected. In the case of multilayer networks, the infection may diffuse over inter- and intra-layer connections at different speeds, meaning that we have different infection rates (i.e., transmissibilities) across the links of each layer and also the links between the layers.
Therefore, most of the works on spreading processes over multilayer networks~\cite{Dickison2012a, Wang2011a, Qian2012, Zhao2014770, Marceau2011, Buono2013, PhysRevE.81.036118, PhysRevE.88.022801, Sanz2014,Min2013,Cozzo2013a,Saumell-Mendiola2012,Zhao2014770} have extended epidemic-like models by considering different infection rates dependent on the types of the layers.

A recent contribution in the context of multiplex networks~\cite{DarabiSahneh2013a} proposed a generalized epidemic mean-field (GEMF) model capable of Modeling epidemic-like spreading processes
  with more complex states in multiplex network layers (compared to
  two or three states in the SIS and SIR models).

\subsubsection{Decision-based Models}\label{sec:mod:dec}

Decision-based models (also called threshold models) are based on the
idea that an agent may decide to adopt a particular behavior depending
on the behavior of its neighbors~\cite{Granovetter1978, Watts2002,
  Centola2007a, Morris2000}. For example, a user can join a
demonstration if a suitable fraction of his/her friends decide to
participate to the event as well. Although \emph{threshold} models may
be the more common name in the physics literature, we use
\emph{decision-based} models~\cite{Easley2010}, to emphasize that
decision is an inherent characteristic of these models.

Existing decision-based studies follow two different approaches~\cite{Easley2010}: (i) \emph{informational} and (ii)
  \emph{direct-benefit} effects.

\textbf{Informational Effects:} In this approach, making decision
  is based on the indirect information about the decisions of
  others. Granovetter presented the first decision-based model, called
  Linear Threshold Model (LTM)~\cite{Granovetter1978}. In LTM, each
  node chooses a threshold value $T_\textit{LTM} \in [0,1]$ and adopts
  a new behavior if and only if at least a fraction $T_\mathit{LTM}$
  of its neighbors has already adopted the new behavior. Based on LTM,
  Watts~\cite{Watts2002} studied the roles of thresholds and network
  structure on information diffusion. The Watts threshold model has
  also been generalized for multiplex networks in~\cite{Brummitt2012a,
    Yagan2012}. All these extensions lead to the conclusion that
  multiplex networks are more likely to produce global adoption
  cascades than monoplex networks.

\textbf{Direct-Benefit Effects:} This approach assumes that there
  are direct payoffs from copying the decisions of others~\cite{Kossinets2003}. Therefore, game-theoretic modeling is at the
  center of this type of decision-based models. In~\cite{Ramezanian2014} the authors generalize the model of networked
  coordination games~\cite{Morris2000} for diffusion processes on
  multiplex networks.  Given a payoff matrix for choosing two possible
  behaviors $A$ and $B$, each node is playing a game with its
  neighbors across all layers. In each round, all nodes update their
  strategies based on the whole payoff (i.e., the sum of all the
  payoffs collected in all layers). In~\cite{Ramezanian2014} the
  authors derive a lower bound for the success of a new behavior,
  defined as the eventual adoption of the new behavior across all
  nodes in the network.

\subsection{Theoretical approaches for analyzing spreading models in multilayer networks}\label{sec:mod:anal}

The dynamics of spreading models have been studied using
  different well established mathematical methods. We now describe
  some of those analytic approaches.

\subsubsection{Generating function}

The generating function technique is widely used in the analysis
  of stochastic processes~\cite{Wilf2006}. Generating functions can
  uniquely determine a discrete sequence of numbers, and can be useful
  for computing probability density functions, moments,
  limit distributions, and solutions of recursions and linked
  differential-difference equations~\cite{Resnick1992}.  Generating
  functions have also been used to study branching and percolation
  processes as two important stochastic processes for modeling spread
  of epidemics over networks.

The branching process model is a simple framework for modeling
  epidemics on a network~\cite{Easley2010}. Suppose that an infected
  agent may come in contact with $k$ other agents while it is
  infectious, and can spread the disease to each of those with
  probability $p$. Each of those $k$ agents (\emph{first wave}) can
  then get in contact with $k$ other agents, so that the disease could
  be spread to $k^2$ individuals (\emph{second wave}), and so on.
  Questions like whether the process dies out after a set of infection
  waves or continues indefinitely are of significant interest in the
  analysis of this process. A theoretical framework for branching
  processes in multiplex networks has been recently
  developed~\cite{Buono2013}.

Branching processes, however, can not be applied in situations
  when the probability to transmit a disease depends on the past
  history of the destination agent, e.g., if it has already been
  infected and become immune as in the SIR
  model~\cite{harris2002}. The steady-state behavior of the SIR model
  can be analyzed by mapping this process into a bond percolation
  process on graphs~\cite{Newman2002, Kenah2007, Hebert-Dufresne2013},
  and then using existing results for graph
  percolation~\cite{Callaway2000}.

Percolation theory studies the structure of connected clusters in
  random graphs. 
 It has been shown that there exists a critical probability
 $p_c$ such that for $p > p_c$ the random graph has a giant connected
 component (GCC). A percolation transition occurs at the critical
 occupation probability $p_c$, which is the point of
 appearance/disappearance of a GCC.
  In~\cite{Baxter2014} the
  authors extends percolation theory to multiplex networks by
  introducing the concept of \emph{weak bootstrap percolation} and
  \emph{weak pruning percolation}. The authors show that these two
  models are distinct and give origin to different critical behaviors
  on the emergence of critical transitions, unlike their equivalence
  in the case of single layer.
  

\subsubsection{Markov-Chain Approximation}

The Microscopic Markov-Chain Approximation (MMA) is an established
approach to study the microscopic behavior of epidemic dynamics, e.g.,
the probability that a given node will be
infected~\cite{Wang03,Chakrabarti2008}. This approach can further be
categorized as (i) Discrete-time version~\cite{GomezSpreadingEPL2010},
and (ii) Continuous-time version~\cite{mieghem2009}. In a
  discrete-time MMA framework, \cite{VidaGC13} 
  study the malware propagation on a multiplex network
  where each node in all layers are in same state however, the diffusion process is totally
  independent on each layer.
  The results show that the dynamics of a SIS
  contagion process in multiplex networks are equivalent to the
  spreading in a single layer which is governed by an effective
  contagion matrix. This allows us to treat epidemic spreading as in a
  single network. 
  The authors observed that coupling of layers helps the viruses propagation.
  Moreover, in~\cite{Wu2014} the authors study epidemic spreading in multiplex
networks by using a combination of discrete-time and continuous-time
MMA approaches. 
More in the context of Markov-Chain approximation, the authors in~\cite{Wei2012,
    CompMemProOnNwPersp} study the spreading of two interacting
  processes in an arbitrary multiplex network by approximating the
  diffusion process as a discrete-time non-linear dynamical system.

\subsubsection{Mean-field theory}

Markovian modeling is a common approach for modeling stochastic
processes between nodes, or in more technical sense, interacting
agents in a network. Unfortunately, large markovian models may become
intractable; mean-field theory studies the behavior of such large and
complex models by considering a simpler model. Instead of computing
the effect of all agents, mean-field theoretic approaches consider a
small averaged effect and an external field, replacing the interaction
of all other agents. Mean-field theory has been used to capture the
macroscopic behavior of the epidemic dynamics such as epidemic
threshold and infection size of epidemic-like
models~\cite{Pastor-Satorras2001}. This theory has been widely applied
to epidemic processes in monoplex networks, under different
assumptions and settings~\cite{Pastor-Satorras2001}. Some recent
works use mean-field approximation for analyzing epidemic-like models
in multilayer networks~\cite{DarabiSahneh2013a, Saumell-Mendiola2012}.

In~\cite{Saumell-Mendiola2012} the authors determine that the SIS
epidemic threshold in an interconnected network with two layers is
smaller than the epidemic thresholds of the two networks separately
even when the epidemics can not propagate on each network separately
and the number of coupling connections is small; the same result may
apply to the SIR model. In~\cite{DarabiSahneh2013a} the authors
  analyze a generalization of the epidemic-like models for multilayer
  networks. Mean-field approximation allows the description of the
  model with a number of nonlinear differential equations with
  linearly growing state space.

\subsubsection{Game theory}

Some researchers have analyzed spreading processes using
  game-theoretical framework in monoplex as well multilayer
  settings. Game theory allows modeling the user's behavior to
  understand the effect of cooperation and competition on information
  dissemination. For example, the model proposed
  in~\cite{Zinoviev2010} explicitly represents feature of each
  spreading agent such as reputation and desire of popularity, in
  addition to the usual structure of the network. The model shows that
  the emergence of social networks can be explained in terms of
  maximization of the game-theoretical payoff.

Similarly, the information diffusion model described
  in~\cite{Qiu2012} takes into consideration various factors
  pertaining to humans, such as knowledge and belief persuasion, and
  shows that the speed of spreading is influenced by the features of
  each individual in the network.

Apart from social networks, studies have also been conducted to
  understand the information propagation in other settings such as
  vehicular networks~\cite{Banerjee2013}.  Recently, game theory has
  also been studied in multilayer settings. For example,
  in~\cite{Ramezanian2014}, the authors have studied the diffusion of
  innovation using the networked coordination game.

%% file: observing.tex
\section{Spreading Dynamics on Multilayer Networks} \label{sec:obsr}

The dynamics of spreading processes, e.g., speed or pattern of
spreading, are influenced by the properties of underlying multilayer
network. In this section we discuss the effect of various properties
considered in the literature for interconnected networks
(Section~\ref{InterNet}) and multiplex networks
(Section~\ref{MultiplexNet}). In Table~\ref{tab:addlabel}, we
summarize and consolidate the discussions.

Aggregating different layers into a single network is one possible way
to study multiplex networks~\cite{DeDomenico2013}.  For example,
in~\cite{Sun2014635} the authors reduce a multilayer network to a
weighted monoplex network, so that the epidemic threshold and
infection size of SIR and SIS models on the multiplex networks can be
studied by looking at the reduced graph. However, disregarding the
inherent multiplex nature of a system could lead to loss of
information and wrong conclusions~\cite{Cozzo2013a}. In this section
we will be focused on work that explicitly considers the multiplex
nature of the systems.

\subsection{Interconnected Networks}\label{InterNet}

The dynamics of different types of diffusion processes in
  interconnected networks can be affected by spectral properties of
  the combinatorial supra-Laplacian of underlying
  graph~\cite{Gomez2013, Radicchi2013, Radicchi2014}. This matrix and
  consequently its properties are strongly affected by inter-layer
  coupling, i.e., coupling (or interaction) strength between
  layers. In particular, \cite{Radicchi2013} shows that changing the
  second eigenvalue of algebraic connectivity of an interconnected
  network has two distinct regimes (layers are decoupled or
  indistinguishable) and a structural transition phase between them.

Most of the works on spreading processes in interconnected
  networks studied the impact of inter-layer connections, in terms of
  \emph{Interaction strength between layers} and \emph{Inter-layer
    pattern}. Next, we review these works.

\subsubsection{Interaction strength between layers}

We start by describing some measures for the interaction
  strength, and mention the works which studied their effect on
  particularly the spreading processes.

\textbf{Second-nearest neighbors: }
The expected number $\kappa$ of neighbors of a node chosen by following an arbitrary link incident to a given source can be computed as $\kappa = \langle k^2 \rangle / \langle
  k \rangle$ where $\langle k^2 \rangle$ and $\langle k \rangle$ are
  the second and first moment of the node degree distribution,
  respectively~\cite{Dorogovtsev2003}. This measure
  considered in~\cite{Dickison2012a} as a measure for coupling
  strength.
In particular, the authors define an interdependent network to be
strongly-coupled if $\kappa_T$ is larger than both $\kappa_A$ and
$\kappa_B$, where $\kappa_A$ and $\kappa_B$ are calculated over the
individual layers $A$ and $B$, and $\kappa_T$ is calculated over the
entire coupled network (i.e., including intra- and inter-layer links).
On the other hand, a network is defined to be weakly-coupled if $\kappa_B
> \kappa_T$ and $\kappa_T > \kappa_A$.  The authors show that in the
case of a spreading disease (modeled by the SIR model) over a
strongly-coupled network, all networks are either in epidemic state or
disease free (with the presence of inter-layer links enhancing
epidemic spreading). However, in the weakly-coupled case a new mixed
phase can exist, with the boundaries dependent on the values of $T$
and $\langle k_{AB} \rangle$, which denote the transmissibility of
the SIR model and the average of inter-layer degrees, respectively. In
this mixed phase, the disease is epidemic on only one layer, and not
in other layers.  Moreover, increasing the inter-layer links only affects
epidemic spreading on the layer with more intra-layer links,
and the epidemic on the layer with lower number of intra-layer links
remain unchanged.

\textbf{Interconnection Topology Measure:} In~\cite{Sahneh2013b},
  the authors propose a purely topological and quantitative measure to
  distinguish strongly-coupled and weakly-coupled cases in an
  arbitrary interconnected networks. For an interconnected network
  with two layers $G_1$ and $G_2$, let $A_{11}$ and $A_{22}$ be the
  corresponding adjacency matrices, and let $A_{12}$ denote the
  connections between the layers. For this network, the coupling
  $\Omega(G_1, G_2)$ of the two layers is computed as $\Omega(G_1,
  G_2)={ {{\alpha}^2 {\Vert A^T_{12}x_1 \Vert}^2_2} \over
    {\lambda_1(A_{11})\lambda_1(A_{22})} }$, where $\alpha$ represents
  the heterogeneity of intra- and inter-layer connections, and $x_1$
  is the eigenvector of $A_{11}$ belonging to $\lambda_1(A_{11})$. In
  this measure, larger $\Omega$ means stronger coupling.

\textbf{Inter-layer Link Density:} Inter-layer link density $d$ can be
defined as the ratio of the existing inter-layer links $m$ between two
layers $A$ and $B$ to the total number of possible such links, giving
$d=m / {(n_A \times n_B)}$, where $n_A$ and $n_B$ are the number of
nodes in layer $A$ and $B$, respectively, of an undirected
interconnected network. The maximal inter-layer link density of a
completely interconnected network is $1$. This interaction strength
measure is used in~\cite{Wang2011a} to study the effects of
inter-layer links on information spreading, modeled with SIR, in
two-layer interconnected networks. The authors find that having more
inter-layer links steadily leads to a much larger infection size. In
addition, their results show that infection peak happens in two
networks at different time, when two networks are sparsely
interconnected and the spreading rate is high enough.

\subsubsection{Inter-layer pattern}\label{subsec:interPatt}

The effect of inter-layer pattern (i.e., how the nodes in different
layers connect to each other) on the dynamics of spreading processes
in interconnected networks has been studied in some recent work.
In~\cite{Parshani2010}, the authors introduce two quantitative
  metrics (called \emph{Inter degree-degree correlation} and
  \emph{Inter-clustering coefficient}) to measure non-random coupling
  pattern between nodes in interconnected networks. Recently, a
  simulation-based study in~\cite{Wang2011a} has shown that the
  inter-layer connections based on the node degree (e.g.,
  interconnections between lowest-degree nodes of the two layers or
  lowest-degree nodes in one layer to highest-degree nodes in other
  layer) have less significant impacts on the infection size than the
  density of interconnections.

Related to this research area, in~\cite{PhysRevE.88.022801} the
authors observe that the epidemic threshold of the SIS model in a
two-layer interconnected network is $1/\lambda_1(M+\alpha N)$ where
the denominator presents the largest eigenvalue of the matrix
$(M+\alpha N)$, $\alpha$ being a real constant for controlling the
infection rate between layers, $M$ being a $2n \times 2n$ matrix
composed of the adjacency matrix of each layer with size $n$, and $N$
being a $2n \times 2n$ matrix that represents the inter-layer links
between layers. Then, they show that $\lambda_1(M+\alpha N)$ tends to
be higher (i.e., a smaller epidemic threshold) if the two nodes $u$
and $v$ with a larger eigenvector component product ${x_u}{y_v}$ are
connected~\cite{PhysRevE.88.022801}.

On similar lines, the effect of correlations between intra-layer and
inter-layer degrees is studied in~\cite{Saumell-Mendiola2012}. In an
interconnected network with two layers $A$ and $B$, let $k_{AA}$ and
$k_{AB}$ (resp. $k_{BB}$ and $k_{BA}$) be the number of a node
intra-layer links and inter-layer links. Then, the correlations
between intra- and inter-layer degrees can be measured by factors
$\langle k_{AA}k_{AB} \rangle$ and $\langle k_{BA}k_{BB} \rangle$. The
authors address three different inter-layer patterns based on this
type of correlation: (i) random coupling, (ii) linear correlations,
and (iii) superlinear correlations. Their results show that if this
correlation is strong enough, the outbreak state may arise even if the
epidemic threshold is not satisfied in any of the two networks
separately.

\subsection{Multiplex Networks}\label{MultiplexNet}

Various topics about spreading processes have been addressed in
  multiplex networks. Here we review some of the most relevant works.

\subsubsection{Intra-layer structure}

Epidemic dynamics depend not only on how the links are distributed
between layers, but also across the same
layer~\cite{Qian2012,Qian2013}. In~\cite{Qian2012}, the authors
address information diffusion in a social-physical multiplex network
where the information could spread between individuals either through
physical or online social networks. They address the effect of clique
structures in physical networks (i.e., groups of people who are close
to each other) on the epidemic threshold and infection size. In their
analytic study, based on heterogeneous bond
percolation~\cite{Soderberg2003a}, they show that in large size
cliques information spreads faster.  To this end, they define three
types of link (i) Type-0 (intra-clique) links in the physical layer,
(ii) Type-1 (inter-clique) links in the physical layer, and (iii)
Type-2 (online) links. Let $d_w$ and $d_f$ denote the numbers of
type-1 and type-2 links of a node, respectively. The, with high
probability there exists an epidemic state in the entire network when
$\sigma = {{{1}\over{2}}(a_{11}+a_{22}+\sqrt{(a_{11} - a_{22})^2+4
    a_{12} a_{21}})} > 1$, where $a_{11}= \mathrm{E}[(d_w)^2]/
\mathrm{E}[d_w] -1$, $a_{12}=\mathrm{E}[d_w d_f]/\mathrm{E}[d_w]$,
$a_{21}=\mathrm{E}[d_w d_f]/\mathrm{E}[d_f]$ and
$a_{22}=\mathrm{E}[(d_f)^2]/\mathrm{E}[d_f]-1$.  They also observe a
sharp increase in the percentage of individuals (14\% to 80\%)
receiving the message when the average clique size increases from 1 to
2. In~\cite{Qian2013}, the same authors show that a larger size online
social network may not lead to outbreak in a social-physical network.

\subsubsection{Layer similarity}

One aspect that may influence the spreading behavior in multiplex
networks is the similarity (or lack of) between layers. There are two
important metrics for measuring the level of inter-layer similarity:
degree-decree correlation and average similarity of neighbors.
\textbf{Degree-degree correlation} describes the correlation of
degrees of nodes in different layers~\cite{Zhao2014770, Marceau2011},
analogously to degree correlation in monoplex
networks~\cite{Newman2002}. This type of correlation can be measured
by factors $\langle k_{AA}k_{BB} \rangle$, where $k_{AA}$ and $k_{BB}$
are the number of a node's intra-layer links in layer $A$ and $B$,
respectively.  \textbf{Average similarity of neighbors} is defined as
$\alpha = {{\sum_iK_C(i)}/ {\sum_i K_A(i)+K_B(i)-K_C(i)}}$, where
$K_A(i)$ (respectively $K_B(i)$) is the number of neighbors of node
$i$ in layer $A$ (respectively $B$), and $K_C(i)$ is the number of
common neighbors of node $i$ in layers $A$ and
$B$. In~\cite{Marceau2011, Zhao2014770}, the authors study the impact
of average similarity between two layers on both epidemic threshold
and infection size. They show that a strong positive degree-degree
correlation of nodes in different layers could lead to a low epidemic
threshold and a relatively smaller infection size. Interestingly,
these measures are not significantly affected by the average
similarity of neighbors.

\subsubsection{Layer-switching cost}\label{sec:switch}
In some systems modeled as multiplex networks, the diffusion of a
process from one layer to another may involve non-zero layer-switching
cost or overhead. For example, retweeting a tweet in Twitter may be
more likely than sharing it over other online media (e.g., Facebook)
because of the additional effort required in switching the
communication channel (cost overhead)~\cite{Min2013}. As another
example, in the transportation network of a city where the same
locations can be part of both subway and bus networks, one can
consider the layer-switching cost to move from the subway lines to the
bus route. This cost can be both financial or can represent the time
required to physically change layers~\cite{DeDomenico2013a}.

The effect of the layer-switching cost on the spreading processes has been studied in~\cite{Min2013, Cozzo2013a}. In a still unpublished report~\cite{Min2013} the authors define layer-switching cost by considering the difference between transmissibilities (i.e., effective infection rates $\lambda$) in the SIR model for intra- and inter-layer links.
They show that the epidemic state will appear if the largest eigenvalue $\Lambda$ of the simplified Jacobian matrix $\mathbf{J}=$
 $\begin{pmatrix}
 T_{11} \gamma_1 &  T_{21} \Gamma_1 \\
 T_{12} \Gamma_2 & T_{22} \gamma_2
 \end{pmatrix}$
is greater than one (i.e., $\Lambda > 1$), where $\gamma_i = (\langle {k_i}^2 \rangle - \langle {k_i} \rangle) / \langle {k_i} \rangle$ ($\langle {k_i} \rangle$ and $\langle {k_i}^2 \rangle$ are the first and second moment of the degree distribution of the layer $i$), $\Gamma_i= \langle k_i k_j \rangle / \langle {k_i} \rangle$, and $T_{ij}$ is the transmissibility over the link between layer $i$ and $j$.
They show that $\Lambda$ is a function of the node degrees $\delta z$ and infection rates $\delta\lambda$, and study their effect on the epidemic threshold.
In particular, when both layers have the same average degree the epidemic threshold increases for larger difference between intra- and inter-layer infection rates as it gets more difficult to spread to other layers (high layer crossing overhead). For the constant difference in rates, if the difference of the average degree of the two layers gets larger (i.e., a layer becomes denser), the epidemic threshold decreases as denser layers facilitate spreading. Finally, they find a threshold for the difference of average degrees, above which the epidemic threshold decreases as the difference in rates becomes larger. These results have been obtained on Erd\H{o}s-Renyi random graphs.

Similar findings were presented in~\cite{Cozzo2013a}, where authors study the SIS model in multiplex networks using a contact-contagion formulation with different infection rates for intra and inter-layers. They observed that the layer with largest eigenvalue controls the epidemic threshold of the entire network.

\begin{table*}[htbp]
\begin{flushleft}
\caption{Summary of the aspects and variables studied in publications regarding spreading processes in multilayer networks.
First column: Topics under study.
Second column: Reference.
Third column: Type of underlying multilayer network; {``I''}: Interconnected network, {``M''}: Multiplex network.
Fourth and Fifth columns: intra- and inter-layer connections between layers in underlying multilayer network; {``-''}: Node sets are identical in different layers, {``Par''}: A fraction of nodes are present in all layers with some probability, {``Syn''}: Synthetic dataset, {``Real''}: Real dataset, {``DD''}: Degree-Driven network, {``SF''}: Scale-Free network, {``ER''}: Erdos-Reyni network, {``SW''}: Small-World network, {``R''}: Random inter-connection, {``Corr''}: Inter-layer connections with different correlation, {``LL''}: Interconnections between lowest-degree nodes of the two layers, {``LH''}: Interconnections between lowest-degree nodes in one layer to highest-degree nodes in other layer, {``HH''}: Interconnections between highest-degree nodes of the two layers.
Sixth column: Spreading models.
Seventh column: Measures for analyzing spreading properties; {``ETh''}: Epidemic Threshold, {``IS''}: Infection Size, {``TE''}: Temporal behavior of the prevalence, {``Sur''}: Survival probability, {``CV''}: Cascade Velocity, {``CS''}: Cascade Size, {``STh''}: Survival threshold, {``DTh''}: Absolute-dominance threshold.
Eighth column: Theoretical Approach; {``Gn''}: Generating function, {``MC''}: Microscopic Markov-Chain approximation, {``MF''}: Mean-Field theory, {``Gm''}: Game Theory.
}
\end{flushleft}
  \footnotesize

    \begin{tabular}{|c|c|c|c|c|c|c|c|}
    \hline
     Topics     & \multirow{2}{*}{Paper} & \multicolumn{3}{c|}{Underlying Multilayer Network}  & \multicolumn{2}{c|}{Spreading Properties} & Theoretical \\ \cline{3-7}
       under study   & \multicolumn{1}{c|}{} & Type  & Intra-layer & Inter-layer  & Model & Measures & Approach \\

     \hline
    \multirow{2}{*}{Interaction strength}
    & \cite{Dickison2012a}    & I     & DD  & DD & SIR  & ETh, Sur & Gn\\
     & \cite{Sahneh2013b}    & I     & DD & DD & SIS  & ETh & MF\\
    & \cite{Wang2011a}      & I     & Real (AS), Syn (ER-AS, SF-AS) & Syn (R, LL, LH, HH)   & SIR     & IS,
TE      &  - \\

    \hline
     \multirow{2}{*}{Inter-layer pattern}
    &  \cite{PhysRevE.88.022801} & I  & DD, Syn (ER-SF) & DD, Syn (R)  & SIS  & ETh   & MF\\
    & \cite{Wang2011a}      & I     & Real (AS), Syn (ER-AS, SF-AS)  & Syn (R, LL, LH, HH) &  SIR     & IS,
TE      & - \\
    &  \cite{Saumell-Mendiola2012} & I   & DD, Syn (ER-ER)  & DD, Syn (R, Corr) & SIS   & ETh, TE  & MF \\

  \hline
     \multirow{1}{*}{Intra-layer Structure}
    &  \cite{Qian2012} & M  & DD, Syn (ER-SF)  & -  &  SIR  & IS, ETh   & Gn\\

   \hline
     \multirow{2}{*}{Inter-layer Similarity}
    &  \cite{Zhao2014770} & M    & DD, Syn (ER-ER, ER-SF, SF-SF)  & -   & SIR   & IS, ETh   & Gn \\
   &  \cite{Marceau2011} & M     & DD  & -    &  SIR   & IS   & MF \\

   \hline
     \multirow{2}{*}{Layer-switching Cost}
    & \cite{Min2013} & M     &  DD, Syn(ER-ER)   & -  &  SIR  & IS, ETh   & MF \\
    & \cite{Cozzo2013a} & M  & DD, Real(Twitter)  & -   & SIR  & ETh  & MC\\

  \hline
    \multirow{1}{*}{Diffusion velocity}
    &  \cite{Li2014}     & M    & Syn(ER-ER, SW-SW) & -  & Watts   & CV, CS  & - \\

    \hline
    \multirow{2}{*}{Partially overlapping}
    &  \cite{Yagan2013}     & M     & CDD, Syn(ER-ER,SF-SF) & Par  & SIR      & IS, ETh     & Gn \\
    &  \cite{Buono2013}  & M     & DD, Syn(ER-ER,SF-SF) & Par   & SIR & IS, ETh     & Gn \\

    \hline
     \multirow{6}{*}{Interacting Processes}
   &  \cite{PhysRevE.81.036118} & M  & DD   & -    & SIR      & ETh  & Gn \\
    &  \cite{Marceau2011} & M     & DD     & -   & SIR   & IS   & MF \\
       &  \cite{Sanz2014} & M     & Syn (ER-ER, SF-SF)   & -    & SIS      & ETh, TE  & MF \\
    &  \cite{Budak2011}  & M     & Real   & -   & ICM   & IS  & - \\
   &  \cite{Wei2012, CompMemProOnNwPersp} & M & Syn (ER-SF), Real  & -    & SI1I2S  & ETh & MC \\
   &  \cite{DarabiSahneh2014} & M & DD, Syn (ER-SF) & -  & SI1I2S  & IS, STh, DTh & MF \\

  \hline
    \multirow{5}{*}{Diffusion of Innovations}
  &  \cite{Brummitt2012a}     & M  & DD, Syn (ER-SF) & -  & Watts  & IS  & Gn \\
  &  \cite{Yagan2012}     & M   & DD, Syn (ER-ER) & -  & Watts   & IS  & Gn \\
  &  \cite{Hu2014a}     & M    & DD, Real & -  & Watts  & IS  & Gn \\
  &  \cite{Ramezanian2014}     & M    & Syn (ER-ER) & - & Game  & IS & Gm\\

  \hline
    \multirow{1}{*}{Effect of resource constraints}
    &  \cite{Shai2012}   & M  & Syn (ER-ER,BA-BA) & -  & SIR  & Eth, IS  & - \\

\hline
    \end{tabular}%
      \label{tab:addlabel}%

\end{table*}%

\subsubsection{Diffusion velocity}\label{sec:diffVel}

The presence of multiple layers can impact the speed at which a piece
of information can diffuse through the network; intuitively, one would
expect that multiple layers speed up the diffusion process since more
links are available and nodes can receive more pieces of information
from multiple communication channels. This intuition has indeed been
confirmed in~\cite{Yagan2013, Shao2011, Gomez2013}: the authors show
that the coupling of two layers in multiplex networks can lead to
speed up a spreading process in the entire network.

However, some empirical studies point out that different link
types~\cite{Brockmann2006, Miritello2011, Karsai2011} and
topologically inefficient paths~\cite{Tang2011} may actually decrease
the spreading speed in monoplex networks; this suggests that this area
of research needs more attention. Recently~\cite{Li2014} addressed the
velocity of the cascade process in multiplex networks by considering
the role of inter-layer links. In this simulation-based study, the
authors find that the obstruction of an inter-layer link
connecting the shortest paths distributed in multiple layers leads to a
slower spreading process in multiplex networks. In this context,
  the results in~\cite{DeDomenico2014} on different types of random
  walks on multiplex networks have important implications for
  spreading processes. The authors show that the time required for a
  random walker to visit the nodes depends on the underlying topology,
  the strengths of inter-layer links and the type of random walk.

\subsubsection{Partially overlapped multiplex networks}\label{sec:par}

In partially overlapped multiplex network, only a fraction of the
nodes are present in all layers~\cite{Zhong2014}.
It has been observed that overlap among the various layers adds robustness to the network~\cite{Cellai2013}. In addition, in an empirical study~\cite{Szell2010}, using a large multilayer real dataset, the authors find out that nodes' behavior might differ in different layers. Recently, to understand the overlap between layers, Bianconi proposed a statistical mechanics framework~\cite{Bianconi2013}.

In~\cite{Yagan2013} the authors study SIR dynamics on a two-layer
multiplex social-physical network.
The first layer represents a physical information network where
information spreads through face-to-face communication or direct phone
calls; the second layer represents a social network. The authors
observe that epidemic diffusion (percolation) can happen in the
cojoint network, even if no percolation happens within each individual
network. Moreover, the authors also find that the fraction of nodes
who receive an information item is significantly larger in the
overlapped network, compared with the case when the networks are
disjoint.

In~\cite{Buono2013} the authors propose a theoretical framework to
study the effect of overlapping on SIR tree-like spreading processes.
They show that the epidemic threshold of a multiplex network with two
layers $A$ and $B$ depends on both the topology of each layer and the
fraction $q$ of nodes present in both the layers. When $q$ approaches
zero then the diffusion process mostly happens in layer $A$, while
when $q$ approaches 1 then the diffusion process happens on the fully
overlapped multiplex network. Assuming that an infection is started
from a randomly chosen node in layer $A$, the authors observe that in
the limit $q \rightarrow 0$ the epidemic threshold of the whole
network is $T_c={{1} / {(\beta_A -1)}}$ where $\beta_A$ is the
branching factor of layer $A$. On the other hand, when $q \rightarrow
1$ the epidemic threshold becomes $T_c={{1} / {\sqrt{[\beta_A -
        \beta_B]^2 + 4 {\langle \beta_A \rangle} {\langle \beta_B
        \rangle}}}}$. This result implies that the presence of shared
nodes lets the epidemic threshold of the layer with the lower
propagating capability affect the threshold of the other layer.

\subsubsection{Interacting spreading processes}\label{InterProc}

In the real world, many spreading processes may happen at the same time
over the same network: for example, multiple diseases may spread
concurrently on the same population and produce different
cascades~\cite{Karrer2011}. These processes may interact with each
other so that the dynamics of one of the diseases may be affected by
those of the others. Moreover, depending on the nature of each diffusion
process, the underlying cascades can differ.

The interaction of different spreading processes on
  monoplex networks can be addressed in the settings of multiplex
  networks. A common assumption is that spreading processes can become
  extinct; in this case, one process will dominate the other one
  even when the diffusion rates of both of them are above the epidemic
  threshold~\cite{Prakash2012}.  Recently,
  in~\cite{VandeBovenkamp2014} the authors relax this assumption and
  address the domination time. They find that it depends on the number
  of infected nodes at the beginning of the domination period.

Other works used a game-theoretical framework to investigate
  interacting spreading processes. \cite{Kostka2008} addressed the
  diffusion of competing rumors in social networks as a strategic game. 
  It has been shown that being the player that starts the game for the 
  rumors is not always an advantage. Compared to this work,
  \cite{Goyal2012,Tzoumas2012,Fazeli2012,Bharathi2007,Dubey2006} have
  studied the competition between companies who use their resources to
  maximize the adoption of their product in a social network. There is
  a subtle difference between these works: \cite{Goyal2012} uses a
  stochastic model, whereas~\cite{Tzoumas2012} uses a deterministic
  model, and in~\cite{Fazeli2012} individuals made rational
  decisions. In~\cite{Bharathi2007} the authors presented a game
  theoretic model based on local influence process, while
  in~\cite{Dubey2006} a local quasi linear model is exploited.

An important step towards a theoretical framework for interacting
processes was taken in~\cite{PhysRevE.81.036118}. Extending the bond
percolation analysis of two virus spreading processes for a two-layer
network~\cite{Newman2005}, the authors addressed the interaction
between two SIR processes spreading successively on a multiplex
network.  They find that cross-immunity (through the interaction
between processes) is more effective where high-degree nodes in
different layers are connected. However, their analytic approach is
static and does not cover the evolution of the system over time. This
issue was considered in~\cite{Marceau2011}, where authors addressed
the interaction between spreading processes on multiplex networks in
terms of the heterogeneity level of contact patterns between nodes,
various degree correlations and overlapped links between the
layers. By considering two interacting processes, the first being an
undesirable disease and another being an immunizing process, the
authors have shown that the positive degree correlation increases the
efficiency of immunization, while overlap facilitates the invasion of
disease. In~\cite{Sanz2014} the authors proposed a framework based on
mean field theory to study the diffusion of two concurrent processes
that allows to derive the epidemic threshold of each
process. Moreover, this approach can be extended to various epidemic
models (such as SIR, SIS, and SEIR). They found that the epidemic
thresholds of both processes depend on the parameters that
characterize the underlying network structure and on the dynamics of
each process.

Some related works have been proposed in the computer science
community. In~\cite{Budak2011}, the authors have studied the problem
of limiting misinformation propagation in a social network, called
influence limitation. They have extended the independent cascade model
(ICM) to analyze the dynamics of multiple cascades over a multiplex
network. Moreover, \cite{Wei2012, CompMemProOnNwPersp} have
  studied the spreading of two interacting memes, modeled as
  $SI_1I_2S$ (an SIS-type model), in an arbitrary two-layer multiplex
  network.  In this model, each node can be infected by virus 1 or 2
  (represented as $I_1$ and $I_2$, respectively).  They show that the
  meme with larger first eigenvalue will eventually prevail in the
  entire networks.  However, this result is challenged
  by~\cite{DarabiSahneh2014} where the authors study the long-term
  coexistence of two $SI_1I_2S$ virus spreading process over an
  arbitrary multiplex network (note that the authors referred this
  model as $SI_1SI_2S$).  They find that the long-term coexistence of
  both viruses depends on the structural properties of the underlying
  multiplex network as well as epidemic-related factors. In
  particular, they show that the negative correlation of network
  layers makes it easier for a virus to survive, but the extinction of
  the other virus is more difficult.

\subsubsection{Diffusion of Innovations}\label{sec:diffIno}

Diffusion of an innovation (new behavior, ideas, technology,
  products) over networks and the role of underlying network in its
  dynamics has received considerable interest in social sciences and
  economics~\cite{Morris2000,Rogers2003,Immorlica2007,MatthewO.Jackson2005}.
  Recently, this problem has been studied in the framework of
  multiplex networks. \cite{Brummitt2012a} studies the condition
  and size of global spreading cascades of innovations in a multiplex
  network with multiple types of interactions by using an extension of
  Watts' threshold model. In particular, they assume that a node
  becomes infected if the fraction of infected neighbors in any link
  type is higher than a given threshold. The authors
  in~\cite{Yagan2012} propose a content-dependent threshold model in
  which each link type is associated with a relative bias in spreading
  a given content (e.g., new product). More in this context, the
  authors of~\cite{Hu2014a} have shown that the existence of a
  multiplex correlated graph is a condition for sustaining a viral
  spreading process. To identify the conditions for viral cascading,
  they map this process to a correlated percolation model. Considering
  the approach of direct-benefit effects, \cite{Ramezanian2014} finds
  a lower bound for the success of an innovation (i.e. how many people
  in the network adopt a specific strategy) in a game-theoretic
  framework.


\subsubsection{Resource constraints}\label{sec:constraints}

In a realistic scenario, nodes of a multiplex network share
  limited resources. This will impact the dynamic of spreading
  processes in such networks; for example, a person shares her/his
  time between his/her accounts in different online social networks
  such as Facebook and Twitter. This is studied in~\cite{Shai2012} by
  using a variation of the SIR model called \emph{constrained SIR}. In
  each step of constrained SIR, there is a maximum value on the number
  of neighbors that each node can infect. The authors find that, in
  agreement with previous studies~\cite{Lee2012}, in the absence of
  resource constraints, positively correlated coupling leads to a
  lower epidemic threshold than a negative correlation. However, in
  the presence of constraints, spreading is less efficient in
  positively correlated coupling than negatively correlated networks.

%% file: application.tex
\section{Applications} \label{sec:app}

Spreading processes in multilayer networks have a large number of
applications, such as understanding the dynamics of
cascades~\cite{Banos2013b,SeedSizeStrAffectCascad}, maximizing the
influence of information in the context of viral
marketing~\cite{Domingos:2001:MNV:502512.502525}, or selecting a
subset of nodes in a network where to place sensors in order to detect
the spreading of a virus or information as quickly as
possible~\cite{Leskovec2007}.

The application areas can be roughly categorized into two classes:
\begin{itemize}
\item \emph{Forward Prediction}: applications that need to steer the
  network into a particular desired state. Virtual marketing and
  influence maximization fall under this category.
\item \emph{Backward Prediction}: applications that require to predict
  how a given piece of information will diffuse in a network.
  Effector/initiator, outbreak detection, cascade detection and
  immunization are some examples under this category.
\end{itemize}

In this section we discuss some applications of information diffusion,
representing the two categories above. 

\subsection{Influence Maximization} \label{sec:infMax}

Influence maximization has the goal of spreading a particular message
as quickly as possible to a large number of nodes. This is usually
done by seeding the information through key ``strategic'' nodes in
such a way that they can help in reaching out most of the network. The
identification of such strategic nodes is therefore essential to
ensure that the message spreads quickly and effectively. The problem
of influence maximization in networks has traditionally been focused
on finding influential nodes, that is, a (possibly small) subset of
nodes that have the maximum influence to spread the
message~\cite{Kempe2005,Chen:2009:EIM:1557019.1557047,Chen:2010:SIM:1835804.1835934}.

Recent works in the context of multilayer networks address the problem
of identifying influential nodes in various domains, such as ranking
scientific authors according to multiple levels of information (e.g.,
citation networks and co-authorship
graphs)~\cite{DiscInfAuthHetAcdNwCoMet}, studying the diffusion of a
virus~\cite{IdnCritNodMulLayNwUnMulVecMalAtk} or identifying the most
active individuals in microblogging platforms based on multiple types
of relationships between individuals~\cite{Li2012a}.

In general, the influential nodes are the top-$k$ nodes according
  to some centrality measure, such as betweenness
  centrality~\cite{Magnani2013a, Sole-Ribalta2014}, eigenvector
  centrality~\cite{Sola2013} or page rank~\cite{Halu2013}. It is
  important to observe that results for monoplex networks do not
  always generalize to multilayer networks; as an example,
  in~\cite{Zhao2014} the authors show that the $k$-shell
  index~\cite{Kitsak2010} proposed for identifying the influential
  nodes in monoplex networks loses its effectiveness in interconnected
  networks, so they introduce a new measure which considers both
  structural and spreading properties.

It is also possible to look at the problem of information
dissemination from a completely different perspective, that is, by
looking at the set of possible messages that can be diffused, and find
which message is likely to survive longer in the network compared to
others. As an example, in~\cite{Wei2012} the authors propose a new
metric to quantitatively assess the probability that a message spreads
more than another; therefore, given a set of different but equivalent
messages, it is possible to select the one which will likely propagate
to a higher fraction of nodes in the multilayer graph.

\subsection{Immunization Strategies}\label{sec:ImmStrat}

How can information dissemination improve the resilience of a
  population against a spreading disease? To answer this question,
  various works have investigated the role of information
  dissemination (or awareness) with respect to the control of a
  disease spreading over multilayer networks. In~\cite{Jo2006} the
  authors consider a two-layer network, where the infection layer
  (where an epidemic spreads) is a Watts-Strogatz small-world network,
  and the prevention layer is modeled as a dynamic process in a
  Barab\'{a}si-Albert scale-free network. The authors observe that, in
  this scenario, epidemic waves are strongly reduced to small
  fluctuations, but in certain situations the prevention layer
  actually helps the disease to survive.
In~\cite{Funk2010a,Funk2009} the authors investigate a SIR model where
better-informed nodes have a reduced susceptibility, showing that this
can raise the threshold for the widespread diffusion of the
infection. In a different kind of study, \cite{ModImpUsrAwrImmStrMosi}
propose the Behavior-Immunity model that allows measurement of
vaccination effect based on the impact of proactive immunization
strategies. In~\cite{Shai2013} the authors study a process in which
SIS dynamics are coupled with a process that rewires intra-layer edges
between susceptible and infected nodes on an interconnected network.

Studies dealing with epidemic diffusion are not only based on
synthetic networks, but consider real networks as well. For example,
in~\cite{LDPM13:exploiting} the authors considered information and
disease spreading processes together, using mobile-phone dataset. Some
researchers have proposed metrics for the control of information
awareness to disease propagation. For example, in~\cite{Granell2013a},
the authors found a meta-critical point for the epidemic onset leading
to disease suppression. This critical point depends on awareness
dynamics and the overlay network structure. An additional study
  from the same authors~\cite{Granell2014} identifies the relation
  between the spreading and immunization processes for a wide range of
  parameters; additionally, in the presence of a \emph{mass-media
    effect} in which most of the individuals are aware of the
  infection, the critical point disappears. Epidemic diffusion in
  two-layer networks (one layer spreading a disease and the other
  diffusing awareness on the infection) is analyzed also
  in~\cite{Bagnoli2014}: the authors conclude that the similarity
  between the two layers allows the infection to be stopped with a
  sufficiently high precaution level.
Interested readers can refer to~\cite{Funk2010c} for additional
references for this research area.

In~\cite{Sahneh2012, Sahneh2014} the authors use the SAIS
  model~\cite{DarabiSahneh2011} to find an optimal infection
  information propagation overlay in an underlying network to improve
  resilience against epidemic spreading. SAIS is an extension of the
  traditional SIS model where 'A' represents a new Alert state, The
  authors prove that the spectral centrality of nodes and links
  determines such overlay network. They find that controlling the
  health status of a small subgroup of the nodes and circulating the
  information has a considerable role in disease prevention. The same
  authors use this model to address the importance of individuals'
  responsiveness in the progress of an epidemic~\cite{Sahneh2012a}.

\subsection{Epidemic Routing in Delay-Tolerant Networking}

Delay Tolerant Networking (DTN)~\cite{dtn} seeks to address the
  issues arising in heterogeneous networks where individual nodes may
  lack continuous connectivity. Many useful types of networks fall
  into this category: for example, a commuter bus equipped with
  short-range communication capabilities can carry messages from one
  stop to another. Other examples include deep space communication,
  where delays can be measured in minutes during which one of the
  endpoints may have moved out of sight, or sensor networks where
  communications must be scheduled at specific points in time to
  preserve power.

Routing and resource discovery on DTNs are more challenging than
  the equivalent problems on regular communication networks, where
  link failures are the exception rather than the norm.
  Traditionally, routing in DTN is achieved using epidemic routing
  algorithms~\cite{dtn-routing} over a (directed or undirected) graph
  whose edges represent the current active links.  Multiple types of
  communication channels may be available at the same time: for
  example, a sensor node could be equipped with both short-range (low
  power consumption, relatively high bandwidth) and long-range (high
  power consumption, low bandwidth) RF links that can be jointly
  described using a two-layer network. Finding the ``best'' route
  according to some latency and energy constraints is an important
  application of forward prediction in multilayer networks.

\subsection{Malware Propagation in the Internet}

Studying the propagation of malware over the Internet, and
  possibly designing networks and applications that can slow down and
  contain malware outbreaks, is an important application of both
  forward and backward prediction in the context of information
  dissemination.

Nodes belonging to modern compute networks include mobile devices
  (smartphones, tabled, portable computers) that are generally
  equipped with multiple wired and/or wireless communication
  interfaces. Moreover, applications interact with other applications
  running on devices that may not be in the immediate
  neighborhood. Therefore, not only the communication channels define
  multiple connection layers, but also the interactions of
  applications should be taken into account as an additional layer.

A piece of malware trying to propagate through the computer
  network may take advantage of all available physical connections to
  spread to other devices, and also hijack applications to infect
  remote nodes. Wang et al.~\cite{Wang2009} studied the diffusion
  dynamics of a mobile phone virus capable of infecting phones by
  bluetooth or through MMS messages. This study is actually an
  interesting example of analysis of malware propagation on a two
  layer network. A link $(u, v)$ between phones $u$ and $v$ exists on
  the first layer if and only if $u$ and $v$ are physically close
  together, so that bluetooth communication is possible.  A link $(u,
  v)$ on the second layer exists if and only if the address book of
  phone $u$ contains the number of phone $v$, so that the malware
  infecting $u$ can try to send a copy of itself to $v$ through MMS.

Obviously, the study can be extended to take into consideration
  other types of links, and therefore additional layers. Understanding
  the spreading pattern of malware over multilayer networks can be
  extremely valuable both for predicting the extension of an infection
  (forward prediction), and also to understand where countermeasures
  can be placed in order to contain the epidemic (backward
  prediction), pretty much in a very similar manner as epidemics among
  living organisms already described in~\ref{sec:ImmStrat}.

%% file: conclusion.tex
\section{Conclusion and open problems}\label{sec:con}

Information diffusion in multilayer networks is an active and not yet
consolidated research field, and therefore offers many unsolved
problems to address. In some cases, phenomena that are quite well
understood in monoplex networks are comparatively not well understood
in the context of multilayer networks; in other cases, completely
novel ideas, algorithms and analysis, specific to multilayer networks
have to be developed. Some research directions are illustrated below.

\noindent\textbf{Empirical study of information diffusion: } In
general, collecting real datasets related to a multilayer network is
non-trivial~\cite{Stopczynski2014}.  This issue is even more
challenging when one tries to gather data on both the diffusion
process and the structure of the underlying multilayer network. To the
best of our knowledge there are no works based on real datasets on
information diffusion in multilayer networks, and the totality of
existing works on multilayer diffusion are based on simulation or
analytic studies.  On the other hand, real-world multilayer networks
are sometimes large and non-trivially observable, since no single
company or institution has full control over all layers.  Network
sampling strategies~\cite{Mehdiabadi2012} can be used to address this
issue by decreasing the expense of processing large real
networks. Thus, it is worth exploring how different sampling
approaches can impact the measurement of diffusion
processes~\cite{Mehdiabadi2012a}. In addition, one can explore if
there are other ways to infer the structure of the diffusion graph,
e.g., by injecting suitable messages at given points and track them
(\emph{graph tomography}).

\noindent\textbf{Metrics and measurements: } Several metrics have been
defined for monoplex networks~\cite{Costa2005}, such as diameter,
distances, and various centrality metrics. Some of these metrics
  have been extended to multilayer networks. For details and recent
  papers in this field, refer to~\cite[Section 4.2]{Kivela2013}
  and~\cite[Section 2.2]{Boccaletti2014}; see also the result of using
  structural metrics for characterizing a real-world multiplex
  network~\cite{Battiston2014}. However, it is important to
investigate if new metrics, specific to multilayer networks, can be
defined. An interesting aspect would be to propose new metrics
specific to time-varying phenomena. Another important research
direction would be to explore how these metrics affect the propagation
of information.

\noindent\textbf{New models for diffusion processes in multilayer
  networks:} As already described in this paper, diffusion processes
in multilayer networks are driven by different mechanisms with respect
to the single layer case. The study of diverse topics such as the
propagation of opinions about a new product over social networks, or
the diffusion of a virus across different species (e.g., avian flu
spreading through birds and humans), requires the development of
suitable diffusion models that take into consideration the existence
and interactions of different layers within a network. In this paper
we have discussed some existing works on diffusion processes in
multilayer networks; other phenomena may require novel diffusion
models to be developed.  For example, the \emph{data mining} approach
proposed recently in~\cite{Sun2012} can be considered for Modeling
information diffusion in heterogeneous information networks.  Another
interesting research direction is modeling and analyzing the
diffusion process on multilayer networks from \emph{game-theoretic}
approach. Information diffusion on monoplex network has already been
studied from game theory perspective, in which authors postulate an
increase in utility for players who adopt the new innovation or learn
the new information if enough of their friends have also
adopted~\cite{Morris2000}.

\noindent\textbf{Data visualization: } An old motto says that ``Seeing
is Believing''. Indeed, many phenomena are first observed, and then
suitable models are built to explain the observations. In the context
of information diffusion, data visualization tools can provide a first
impression of what is going on, and suggest that something worth
investigating may be happening indeed. Information diffusion is a
dynamic phenomenon, requiring an additional dimension (time) to be
visualized~\cite{Beck2014}. Diffusion processes in multilayer networks
also require the visualization of different layers, and it is not yet
clear what is the most effective and understandable way to provide
this kind of information. As a recent contribution in this
  direction, in~\cite{de2014multilayer} the authors introduce a
  methodology for the analysis and visualization of multilayer
  networks implemented in an open-source software called
  \emph{muxViz}.

\noindent\textbf{Time-varying networks: } Many real-world
  networks exhibit a mutable structure, meaning that nodes and links
  change over time~\cite{Holme2012}. The spreading processes on such
  time-varying (monoplex) networks is addressed in recent
  works~\cite{Vazquez2007, Volz2009, Taylor2012, Gauvin2013}. Indeed,
  both types of dynamics (i.e., dynamics of spreading processes and
  dynamics of underlying networks) are considered in this field.
  However, studying this problem in time-varying multilayer networks
  is more difficult~\cite{DBLP:journals/socnet/SnijdersLT13}.
  Recently, in a still unpublished report~\cite{Valdano2014} the
  authors utilize the mathematical formulation of multilayer networks
  proposed in~\cite{DeDomenico2013} to study spreading processes on
  time-varying networks.

\noindent\textbf{ Evolution of underlying network structure and
    spreading process:} The coevolution of spreading processes and
  underlying structures in adaptive (monoplex) networks network, where
  nodes change their neighborhood as a response to receiving new
  information, have been considered in~\cite{Gross2006, Gross2008,
    Marceau2011}. An interesting observation is that changing the
  underlying network, e.g., by reducing or modifying contacts to
  prevent infection, does not always lead to reduction of
  spreading~\cite{Meloni2011}. This problem becomes more complex if
  the underlying network is modeled as a multilayer network,
  requiring further research.

\noindent\textbf{Outbreak detection:} Outbreak detection is a
technique for the detection of spreading of a virus (or information)
in a network as quickly as possible~\cite{Leskovec2007}. The problem
of outbreak detection is worth exploring in the area of multilayer
networks.

%% file: main.bbl
\begin{thebibliography}{100}
\providecommand{\url}[1]{#1}
\csname url@samestyle\endcsname
\providecommand{\newblock}{\relax}
\providecommand{\bibinfo}[2]{#2}
\providecommand{\BIBentrySTDinterwordspacing}{\spaceskip=0pt\relax}
\providecommand{\BIBentryALTinterwordstretchfactor}{4}
\providecommand{\BIBentryALTinterwordspacing}{\spaceskip=\fontdimen2\font plus
\BIBentryALTinterwordstretchfactor\fontdimen3\font minus
  \fontdimen4\font\relax}
\providecommand{\BIBforeignlanguage}[2]{{%
\expandafter\ifx\csname l@#1\endcsname\relax
\typeout{** WARNING: IEEEtran.bst: No hyphenation pattern has been}%
\typeout{** loaded for the language `#1'. Using the pattern for}%
\typeout{** the default language instead.}%
\else
\language=\csname l@#1\endcsname
\fi
#2}}
\providecommand{\BIBdecl}{\relax}
\BIBdecl

\bibitem{Bailley1975}
\BIBentryALTinterwordspacing
N.~T.~J. Bailley, \emph{\BIBforeignlanguage{not specified}{{The mathematical
  theory of infectious diseases and its applications.}}}\hskip 1em plus 0.5em
  minus 0.4em\relax London: Charles Griffin \& Company Ltd, 5a Crendon Street,
  High Wycombe, Bucks HP13 6LE., 1975. [Online]. Available:
  \url{http://www.cabdirect.org/abstracts/19762902036.html;jsessionid=5A9B180AE9D952A504D4D1CDB7FCB0EB?freeview=true}
\BIBentrySTDinterwordspacing

\bibitem{Anderson1992}
\BIBentryALTinterwordspacing
R.~M. Anderson and R.~M. May, \emph{{Infectious Diseases of Humans: Dynamics
  and Control}}.\hskip 1em plus 0.5em minus 0.4em\relax Oxford University
  Press, 1992. [Online]. Available:
  \url{http://books.google.it/books/about/Infectious\_Diseases\_of\_Humans.html?id=wJlTngEACAAJ\&pgis=1}
\BIBentrySTDinterwordspacing

\bibitem{Pastor-Satorras2001}
\BIBentryALTinterwordspacing
R.~Pastor-Satorras and A.~Vespignani, ``{Epidemic Spreading in Scale-Free
  Networks},'' \emph{Physical Review Letters}, vol.~86, no.~14, pp. 3200--3203,
  Apr. 2001. [Online]. Available:
  \url{http://link.aps.org/doi/10.1103/PhysRevLett.86.3200}
\BIBentrySTDinterwordspacing

\bibitem{Moreno2002}
\BIBentryALTinterwordspacing
Y.~Moreno, R.~Pastor-Satorras, and A.~Vespignani, ``{Epidemic outbreaks in
  complex heterogeneous networks},'' \emph{The European Physical Journal B},
  vol.~26, no.~4, pp. 521--529, Apr. 2002. [Online]. Available:
  \url{http://www.springerlink.com/index/10.1140/epjb/e20020122}
\BIBentrySTDinterwordspacing

\bibitem{Newman2002}
M.~Newman, ``{Spread of epidemic disease on networks},'' \emph{Physical review.
  E, Statistical, nonlinear, and soft matter physics}, vol.~66, no. 1 Pt 2,
  2002.

\bibitem{Chakrabarti2008}
\BIBentryALTinterwordspacing
D.~Chakrabarti, Y.~Wang, C.~Wang, J.~Leskovec, and C.~Faloutsos, ``{Epidemic
  thresholds in real networks},'' \emph{ACM Transactions on Information and
  System Security}, vol.~10, no.~4, pp. 1--26, Jan. 2008. [Online]. Available:
  \url{http://dl.acm.org/citation.cfm?id=1284680.1284681}
\BIBentrySTDinterwordspacing

\bibitem{mieghem2009}
\BIBentryALTinterwordspacing
P.~{Van Mieghem}, J.~Omic, and R.~Kooij, ``{Virus Spread in Networks},''
  \emph{IEEE/ACM Transactions on Networking}, vol.~17, no.~1, pp. 1--14, Feb.
  2009. [Online]. Available: \url{http://dx.doi.org/10.1109/tnet.2008.925623}
\BIBentrySTDinterwordspacing

\bibitem{Kleinberg2007}
\BIBentryALTinterwordspacing
J.~Kleinberg, ``{Computing: the wireless epidemic.}'' \emph{Nature}, vol. 449,
  no. 7160, pp. 287--8, Sep. 2007. [Online]. Available:
  \url{http://dx.doi.org/10.1038/449287a}
\BIBentrySTDinterwordspacing

\bibitem{Wang2009}
\BIBentryALTinterwordspacing
P.~Wang, M.~C. Gonz\'{a}lez, C.~A. Hidalgo, and A.-L. Barab\'{a}si,
  ``{Understanding the spreading patterns of mobile phone viruses.}''
  \emph{Science (New York, N.Y.)}, vol. 324, no. 5930, pp. 1071--6, May 2009.
  [Online]. Available:
  \url{http://www.sciencemag.org/content/324/5930/1071.abstract}
\BIBentrySTDinterwordspacing

\bibitem{Meisel2010}
\BIBentryALTinterwordspacing
M.~Meisel, V.~Pappas, and L.~Zhang, ``{A taxonomy of biologically inspired
  research in computer networking},'' \emph{Computer Networks}, vol.~54, no.~6,
  pp. 901--916, Apr. 2010. [Online]. Available:
  \url{http://dl.acm.org/citation.cfm?id=1752615.1752711}
\BIBentrySTDinterwordspacing

\bibitem{GuilleSurveyInfoDiss}
\BIBentryALTinterwordspacing
A.~Guille, H.~Hacid, C.~Favre, and D.~A. Zighed, ``{Information Diffusion in
  Online Social Networks: A Survey},'' \emph{SIGMOD Rec.}, vol.~42, no.~2, pp.
  17--28, Jul. 2013. [Online]. Available:
  \url{http://doi.acm.org/10.1145/2503792.2503797}
\BIBentrySTDinterwordspacing

\bibitem{Borge-Holthoefer2013a}
J.~Borge-Holthoefer, R.~A. Banos, S.~Gonzalez-Bailon, and Y.~Moreno,
  ``{Cascading behaviour in complex socio-technical networks},'' \emph{Journal
  of Complex Networks}, vol.~1, no.~1, pp. 3--24, Apr. 2013.

\bibitem{Pastor-Satorras2014}
\BIBentryALTinterwordspacing
R.~Pastor-Satorras, C.~Castellano, P.~{Van Mieghem}, and A.~Vespignani,
  ``{Epidemic processes in complex networks},'' p.~61, Aug. 2014. [Online].
  Available: \url{http://arxiv.org/abs/1408.2701}
\BIBentrySTDinterwordspacing

\bibitem{PhysRevE.88.022801}
\BIBentryALTinterwordspacing
H.~Wang, Q.~Li, G.~D'Agostino, S.~Havlin, H.~E. Stanley, and P.~{Van Mieghem},
  ``{Effect of the interconnected network structure on the epidemic
  threshold},'' \emph{Phys. Rev. E}, vol.~88, no.~2, Aug. 2013. [Online].
  Available: \url{http://link.aps.org/doi/10.1103/PhysRevE.88.022801}
\BIBentrySTDinterwordspacing

\bibitem{Lee2014}
\BIBentryALTinterwordspacing
K.-M. Lee, J.~Y. Kim, S.~Lee, and K.-I. Goh, ``{Multiplex Networks},'' in
  \emph{Networks of Networks: The Last Frontier of Complexity}, 2014, pp.
  53--72. [Online]. Available:
  \url{http://scholar.google.it/citations?view\_op=view\_citation\&hl=en\&user=Af5XPeYAAAAJ\&citation\_for\_view=Af5XPeYAAAAJ:W7OEmFMy1HYC}
\BIBentrySTDinterwordspacing

\bibitem{Buldyrev2010}
\BIBentryALTinterwordspacing
S.~V. Buldyrev, R.~Parshani, G.~Paul, H.~E. Stanley, and S.~Havlin,
  ``{Catastrophic cascade of failures in interdependent networks.}''
  \emph{Nature}, vol. 464, no. 7291, pp. 1025--8, Apr. 2010. [Online].
  Available: \url{http://dx.doi.org/10.1038/nature08932}
\BIBentrySTDinterwordspacing

\bibitem{Danziger2014}
\BIBentryALTinterwordspacing
M.~M. Danziger, A.~Bashan, Y.~Berezin, L.~M. Shekhtman, and S.~Havlin, ``{An
  Introduction to Interdependent Networks},'' in \emph{Nonlinear Dynamics of
  Electronic Systems}, ser. Communications in Computer and Information Science,
  V.~M. Mladenov and P.~C. Ivanov, Eds.\hskip 1em plus 0.5em minus 0.4em\relax
  Cham: Springer International Publishing, 2014, vol. 438, pp. 189--202.
  [Online]. Available: \url{http://link.springer.com/10.1007/978-3-319-08672-9}
\BIBentrySTDinterwordspacing

\bibitem{Kenett2014}
\BIBentryALTinterwordspacing
D.~Y. Kenett, J.~Gao, X.~Huang, S.~Shao, I.~Vodenska, S.~V. Buldyrev, G.~Paul,
  H.~E. Stanley, and S.~Havlin, ``{Network of Interdependent Networks: Overview
  of Theory and Applications},'' in \emph{Networks of Networks: The Last
  Frontier of Complexity}, ser. Understanding Complex Systems, G.~D'Agostino
  and A.~Scala, Eds.\hskip 1em plus 0.5em minus 0.4em\relax Cham: Springer
  International Publishing, 2014, pp. 3--36. [Online]. Available:
  \url{http://link.springer.com/10.1007/978-3-319-03518-5}
\BIBentrySTDinterwordspacing

\bibitem{Berlingerio2012}
\BIBentryALTinterwordspacing
M.~Berlingerio, M.~Coscia, F.~Giannotti, A.~Monreale, and D.~Pedreschi,
  ``{Multidimensional networks: foundations of structural analysis},''
  \emph{World Wide Web}, vol.~16, no. 5-6, pp. 567--593, Oct. 2012. [Online].
  Available: \url{http://link.springer.com/10.1007/s11280-012-0190-4}
\BIBentrySTDinterwordspacing

\bibitem{Magnani2013}
\BIBentryALTinterwordspacing
M.~Magnani and L.~Rossi, ``{Formation of Multiple Networks},'' in \emph{Social
  Computing, Behavioral-Cultural Modeling and Prediction}, ser. Lecture Notes
  in Computer Science, A.~M. Greenberg, W.~G. Kennedy, and N.~D. Bos,
  Eds.\hskip 1em plus 0.5em minus 0.4em\relax Berlin, Heidelberg: Springer
  Berlin Heidelberg, 2013, vol. 7812, pp. 257--264. [Online]. Available:
  \url{http://dblp.uni-trier.de/db/conf/sbp/sbp2013.html\#Magnani013a}
\BIBentrySTDinterwordspacing

\bibitem{Mucha2010}
\BIBentryALTinterwordspacing
P.~J. Mucha and M.~a. Porter, ``{Communities in multislice voting networks},''
  \emph{Chaos: An Interdisciplinary Journal of Nonlinear Science}, vol.~20,
  no.~4, 2010. [Online]. Available:
  \url{http://scitation.aip.org/content/aip/journal/chaos/20/4/10.1063/1.3518696}
\BIBentrySTDinterwordspacing

\bibitem{Wasserman1994}
\BIBentryALTinterwordspacing
S.~Wasserman and K.~Faust, \emph{{Social Network Analysis: Methods and
  Applications}}, ser. Structural analysis in the social sciences, 8.\hskip 1em
  plus 0.5em minus 0.4em\relax Cambridge University Press, 1994, vol.~8, no.~1.
  [Online]. Available: \url{http://www.amazon.com/dp/0521387078}
\BIBentrySTDinterwordspacing

\bibitem{Wang2013a}
\BIBentryALTinterwordspacing
P.~Wang, G.~Robins, P.~Pattison, and E.~Lazega, ``{Exponential random graph
  models for multilevel networks},'' \emph{Social Networks}, vol.~35, no.~1,
  pp. 96--115, Jan. 2013. [Online]. Available:
  \url{http://www.sciencedirect.com/science/article/pii/S0378873313000051}
\BIBentrySTDinterwordspacing

\bibitem{D'Agostino2014}
\BIBentryALTinterwordspacing
G.~D'Agostino and A.~Scala, Eds., \emph{{Networks of Networks: The Last
  Frontier of Complexity}}.\hskip 1em plus 0.5em minus 0.4em\relax Springer
  Berlin / Heidelberg, 2014. [Online]. Available:
  \url{http://www.springer.com/physics/complexity/book/978-3-319-03517-8}
\BIBentrySTDinterwordspacing

\bibitem{Gao2014}
\BIBentryALTinterwordspacing
J.~Gao, D.~Li, and S.~Havlin, ``\BIBforeignlanguage{en}{{From a single network
  to a network of networks}},'' \emph{\BIBforeignlanguage{en}{National Science
  Review}}, pp. 1--11, Jul. 2014. [Online]. Available:
  \url{http://nsr.oxfordjournals.org/content/early/2014/07/16/nsr.nwu020.full}
\BIBentrySTDinterwordspacing

\bibitem{Roethlisberger1939}
\BIBentryALTinterwordspacing
F.~J. Roethlisberger and W.~J. Dickson, \emph{{Management and the
  Worker}}.\hskip 1em plus 0.5em minus 0.4em\relax Cambridge University Press,
  1939. [Online]. Available:
  \url{http://books.google.com/books?id=z5BZ72cwZlYC\&pgis=1}
\BIBentrySTDinterwordspacing

\bibitem{Gluckman1955}
M.~Gluckman, \emph{{The judicial process among the Barotse of Northern Rhodesia
  /}}.\hskip 1em plus 0.5em minus 0.4em\relax Manchester :: Manchester
  University Press on behalf of the Rhodes-Livingstone Institute, Northern
  Rhodesia,, 1955.

\bibitem{Mitchell1969}
\BIBentryALTinterwordspacing
J.~C. Mitchell, \emph{{Social Networks in Urban Situations: Analyses of
  Personal Relationships in Central African Towns}}.\hskip 1em plus 0.5em minus
  0.4em\relax Manchester University Press, 1969. [Online]. Available:
  \url{http://books.google.it/books/about/Social\_Networks\_in\_Urban\_Situations.html?id=8RrpAAAAIAAJ\&pgis=1}
\BIBentrySTDinterwordspacing

\bibitem{Verbrugge1979}
\BIBentryALTinterwordspacing
L.~M. Verbrugge, ``{Multiplexity in Adult Friendships},'' \emph{Social Forces},
  vol.~57, no.~4, pp. 1286--1309, Jun. 1979. [Online]. Available:
  \url{http://sf.oxfordjournals.org/content/57/4/1286.abstract}
\BIBentrySTDinterwordspacing

\bibitem{Breiger1986}
\BIBentryALTinterwordspacing
R.~L. Breiger and P.~E. Pattison, ``{Cumulated social roles: The duality of
  persons and their algebras},'' \emph{Social Networks}, vol.~8, no.~3, pp.
  215--256, Sep. 1986. [Online]. Available:
  \url{http://www.sciencedirect.com/science/article/pii/0378873386900067}
\BIBentrySTDinterwordspacing

\bibitem{Pattison1999}
\BIBentryALTinterwordspacing
P.~Pattison and S.~Wasserman, ``{Logit models and logistic regressions for
  social networks: II. Multivariate relations.}'' \emph{The British journal of
  mathematical and statistical psychology}, vol. 52 ( Pt 2), pp. 169--93, Nov.
  1999. [Online]. Available: \url{http://www.ncbi.nlm.nih.gov/pubmed/10613111}
\BIBentrySTDinterwordspacing

\bibitem{Kivela2013}
M.~Kivel\"{a}, A.~Arenas, M.~Barthelemy, J.~P. Gleeson, Y.~Moreno, and M.~A.
  Porter, ``{Multilayer Networks},'' \emph{Journal of Complex Networks}, pp.
  1--69, Sep. 2014.

\bibitem{Boccaletti2014}
\BIBentryALTinterwordspacing
S.~Boccaletti, G.~Bianconi, R.~Criado, C.~del Genio, J.~G\'{o}mez-Garde\~{n}es,
  M.~Romance, I.~Sendi\~{n}a Nadal, Z.~Wang, and M.~Zanin, ``{The structure and
  dynamics of multilayer networks},'' \emph{Physics Reports}, Jul. 2014.
  [Online]. Available:
  \url{http://www.sciencedirect.com/science/article/pii/S0370157314002105}
\BIBentrySTDinterwordspacing

\bibitem{Shao2011}
\BIBentryALTinterwordspacing
J.~Shao, S.~V. Buldyrev, S.~Havlin, and H.~E. Stanley, ``{Cascade of failures
  in coupled network systems with multiple support-dependence relations},''
  \emph{Physical Review E}, vol.~83, no.~3, p. 036116, Mar. 2011. [Online].
  Available: \url{http://link.aps.org/doi/10.1103/PhysRevE.83.036116}
\BIBentrySTDinterwordspacing

\bibitem{Brummitt2012}
C.~D. Brummitt, R.~M. D'Souza, and E.~a. Leicht, ``{Suppressing cascades of
  load in interdependent networks.}'' \emph{Proceedings of the National Academy
  of Sciences of the United States of America}, vol. 109, no.~12, pp. E680--9,
  Mar. 2012.

\bibitem{Shin2014}
\BIBentryALTinterwordspacing
D.-H. Shin, D.~Qian, and J.~Zhang, ``{Cascading effects in interdependent
  networks},'' \emph{IEEE Network}, vol.~28, no.~4, pp. 82--87, Jul. 2014.
  [Online]. Available:
  \url{http://ieeexplore.ieee.org/lpdocs/epic03/wrapper.htm?arnumber=6863136}
\BIBentrySTDinterwordspacing

\bibitem{gomezPRE2012}
J.~G\'{o}mez-Garde\~{n}es, C.~Gracia-L\'{a}zaro, L.~M. Flor\'{\i}a, and
  Y.~Moreno, ``{Evolutionary dynamics on interdependent populations},''
  \emph{Phys. Rev. E}, vol.~86, no.~2, p. 056113, 2012.

\bibitem{WangSP13}
\BIBentryALTinterwordspacing
Z.~Wang, A.~Szolnoki, and M.~Perc, ``{Optimal interdependence between networks
  for the evolution of cooperation.}'' \emph{Scientific reports}, vol.~3, p.
  2470, Jan. 2013. [Online]. Available:
  \url{http://www.pubmedcentral.nih.gov/articlerender.fcgi?artid=3747507\&tool=pmcentrez\&rendertype=abstract}
\BIBentrySTDinterwordspacing

\bibitem{Jiang2013}
\BIBentryALTinterwordspacing
L.-L. Jiang and M.~Perc, ``{Spreading of cooperative behaviour across
  interdependent groups.}'' \emph{Scientific reports}, vol.~3, p. 2483, Jan.
  2013. [Online]. Available:
  \url{http://www.pubmedcentral.nih.gov/articlerender.fcgi?artid=3748424\&tool=pmcentrez\&rendertype=abstract}
\BIBentrySTDinterwordspacing

\bibitem{Santos2014}
\BIBentryALTinterwordspacing
M.~Santos, S.~Dorogovtsev, and J.~Mendes, ``{Biased imitation in coupled
  evolutionary games in interdependent networks},'' \emph{Scientific Reports},
  vol.~1, pp. 1--19, 2014. [Online]. Available:
  \url{http://www.nature.com/srep/2014/140324/srep04436/full/srep04436.html}
\BIBentrySTDinterwordspacing

\bibitem{Barreto2008}
\BIBentryALTinterwordspacing
E.~Barreto, B.~Hunt, E.~Ott, and P.~So, ``{Synchronization in networks of
  networks: the onset of coherent collective behavior in systems of interacting
  populations of heterogeneous oscillators.}'' \emph{Physical Review E},
  vol.~77, no.~3, p. 036107, Mar. 2008. [Online]. Available:
  \url{http://www.pubmedcentral.nih.gov/articlerender.fcgi?artid=2453534\&tool=pmcentrez\&rendertype=abstract}
\BIBentrySTDinterwordspacing

\bibitem{Bogojeska2013}
I.~M. {A. Bogojeska S. Filiposka} and L.~Kocarev, ``{Observing Dynamical
  Processes in Multiplex Networks by Using Edge Correlation},''
  \emph{Int.J.Comp.Syst.Sci.}, no.~3, pp. 107--112, 2013.

\bibitem{Aguirre2014}
\BIBentryALTinterwordspacing
J.~Aguirre, R.~Sevilla-Escoboza, R.~Guti\'{e}rrez, D.~Papo, and J.~Buld\'{u},
  ``{Synchronization of Interconnected Networks: The Role of Connector
  Nodes},'' \emph{Physical Review Letters}, vol. 112, no.~24, p. 248701, Jun.
  2014. [Online]. Available:
  \url{http://link.aps.org/doi/10.1103/PhysRevLett.112.248701}
\BIBentrySTDinterwordspacing

\bibitem{Easley2010}
\BIBentryALTinterwordspacing
D.~Easley and J.~Kleinberg, \emph{{Networks, Crowds, and Markets: Reasoning
  About a Highly Connected World}}.\hskip 1em plus 0.5em minus 0.4em\relax
  Cambridge University Press, 2010. [Online]. Available:
  \url{http://www.amazon.com/Networks-Crowds-Markets-Reasoning-Connected/dp/0521195330}
\BIBentrySTDinterwordspacing

\bibitem{DeDomenico2013}
\BIBentryALTinterwordspacing
M.~{De Domenico}, A.~Sol\'{e}-Ribalta, E.~Cozzo, M.~Kivel\"{a}, Y.~Moreno,
  M.~Porter, S.~G\'{o}mez, and A.~Arenas, ``{Mathematical Formulation of
  Multilayer Networks},'' \emph{Physical Review X}, vol.~3, no.~4, Dec. 2013.
  [Online]. Available: \url{http://link.aps.org/doi/10.1103/PhysRevX.3.041022}
\BIBentrySTDinterwordspacing

\bibitem{Dickison2012a}
\BIBentryALTinterwordspacing
M.~Dickison, S.~Havlin, and H.~E. Stanley, ``{Epidemics on interconnected
  networks},'' \emph{Physical Review E}, vol. 066109, no.~6, pp. 1--6, 2012.
  [Online]. Available: \url{http://arxiv.org/abs/1201.6339}
\BIBentrySTDinterwordspacing

\bibitem{Eslami2011}
M.~Eslami, H.~R. Rabiee, and M.~Salehi, ``{DNE: A Method for Extracting
  Cascaded Diffusion Networks from Social Networks},'' in \emph{IEEE Conference
  on Social Computing}.\hskip 1em plus 0.5em minus 0.4em\relax IEEE, 2011, pp.
  41--48.

\bibitem{Banos2013b}
\BIBentryALTinterwordspacing
R.~Ba\~{n}os, J.~Borge-Holthoefer, and Y.~Moreno, ``{The role of hidden
  influentials in the diffusion of online information cascades},'' \emph{EPJ
  Data Science}, pp. 1--21, 2013. [Online]. Available:
  \url{http://link.springer.com/article/10.1140/epjds18}
\BIBentrySTDinterwordspacing

\bibitem{Li2014}
Z.~L. Li and J.~Yichuan, ``{Cross-Layers Cascade in Multiplex Networks},'' in
  \emph{International Conference on Autonomous Agents and Multiagent Systems
  (AAMAS)}, Paris, 2014, pp. 269--276.

\bibitem{Min2013}
\BIBentryALTinterwordspacing
B.~Min and K.~Goh, ``{Layer-crossing overhead and information spreading in
  multiplex social networks},'' \emph{arXiv preprint arXiv:1307.2967}, pp.
  1--5, 2013. [Online]. Available: \url{http://arxiv.org/abs/1307.2967}
\BIBentrySTDinterwordspacing

\bibitem{DBLP:conf/asonam/MagnaniR11}
M.~Magnani and L.~Rossi, ``{The ML-Model for Multi-layer Social Networks},'' in
  \emph{ASONAM}.\hskip 1em plus 0.5em minus 0.4em\relax IEEE Computer Society,
  2011, pp. 5--12.

\bibitem{Mehdiabadi2012}
\BIBentryALTinterwordspacing
M.~E. Mehdiabadi, H.~R. Rabiee, and M.~Salehi, ``{Sampling from Diffusion
  Networks},'' in \emph{International Conference on Social Informatics
  (SocialInformatics)}, 2012, pp. 106--112. [Online]. Available:
  \url{http://dblp.uni-trier.de/db/conf/socialinformatics/socialinformatics2012.html\#MehdiabadiRS12}
\BIBentrySTDinterwordspacing

\bibitem{Mehdiabadi2012a}
\BIBentryALTinterwordspacing
------, ``{Diffusion-Aware Sampling and Estimation in Information Diffusion
  Networks},'' in \emph{International Conference on Privacy, Security, Risk and
  Trust and 2012 International Confernece on Social Computing}.\hskip 1em plus
  0.5em minus 0.4em\relax IEEE, Sep. 2012, pp. 176--183. [Online]. Available:
  \url{http://dblp.uni-trier.de/db/conf/socialcom/socialcom2012.html\#MehdiabadiRS12}
\BIBentrySTDinterwordspacing

\bibitem{Leskovec07blogs}
J.~Leskovec, M.~McGlohon, C.~Faloutsos, N.~Glance, and M.~Hurst, ``{Cascading
  Behavior in Large Blog Graphs},'' Apr. 2007.

\bibitem{Magnani2010a}
M.~Magnani, M.~Danilo, and L.~Rossi, ``{Information propagation analysis in a
  social network site},'' in \emph{IEEE International conference on Social
  Network Analysis and Mining (ASONAM)}.\hskip 1em plus 0.5em minus 0.4em\relax
  IEEE Computer Society, 2010.

\bibitem{WangWWW2011}
D.~Wang, Z.~Wen, H.~Tong, C.~Y. Lin, C.~Song, and A.~L. Barab\'{a}si,
  ``{Information spreading in context},'' in \emph{International conference on
  World wide web}, ser. WWW '11.\hskip 1em plus 0.5em minus 0.4em\relax New
  York, NY, USA: ACM, 2011, pp. 735--744.

\bibitem{BakshyWWW2012}
E.~Bakshy, I.~Rosenn, C.~Marlow, and L.~Adamic, ``{The Role of Social Networks
  in Information Diffusion},'' in \emph{International Conference on World Wide
  Web}, ser. WWW '12.\hskip 1em plus 0.5em minus 0.4em\relax New York, NY, USA:
  ACM, 2012, pp. 519--528.

\bibitem{magnani2013factors}
M.~Magnani, D.~Montesi, and L.~Rossi, ``{Factors Enabling Information
  Propagation in a Social Network Site},'' in \emph{The Influence of Technology
  on Social Network Analysis and Mining}.\hskip 1em plus 0.5em minus
  0.4em\relax Springer Vienna, 2013, pp. 411--426.

\bibitem{Gjoka2011}
\BIBentryALTinterwordspacing
M.~Gjoka and C.~Butts, ``{Multigraph sampling of online social networks},''
  \emph{Selected Areas in \ldots}, 2011. [Online]. Available:
  \url{http://ieeexplore.ieee.org/xpls/abs\_all.jsp?arnumber=6027869}
\BIBentrySTDinterwordspacing

\bibitem{epstein08}
J.~M. Epstein, ``{Why Model?}'' \emph{Journal of Artificial Societies and
  Social Simulation}, vol.~11, no.~4, p.~12, 2008.

\bibitem{Cozzo2013a}
E.~Cozzo, R.~A. Ba, S.~Meloni, and Y.~Moreno, ``{Contact-based Social Contagion
  in Multiplex Networks},'' \emph{Physical Review E}, vol.~88, no.~5, p.
  050801, 2013.

\bibitem{Buono2013}
\BIBentryALTinterwordspacing
C.~Buono, L.~G. Alvarez-Zuzek, P.~A. Macri, and L.~A. Braunstein, ``{Epidemics
  in partially overlapped multiplex networks},'' \emph{PloS one}, vol.~9,
  no.~3, p.~5, 2014. [Online]. Available: \url{http://arxiv.org/abs/1310.1939}
\BIBentrySTDinterwordspacing

\bibitem{Qian2012}
D.~Qian, O.~Yagan, L.~Yang, and J.~Zhang, ``{Diffusion of real-time information
  in social-physical networks},'' in \emph{IEEE GLOBECOM}, 2012, pp.
  2072--2077.

\bibitem{Zhao2014770}
\BIBentryALTinterwordspacing
D.~wei Zhao, L.~Li, H.~Peng, Q.~Luo, and Y.~Yang, ``{Multiple routes
  transmitted epidemics on multiplex networks},'' \emph{Physics Letters A},
  vol. 378, no.~10, pp. 770--776, 2014. [Online]. Available:
  \url{http://www.sciencedirect.com/science/article/pii/S0375960114000607}
\BIBentrySTDinterwordspacing

\bibitem{VidaGC13}
R.~Vida, J.~Galeano, and S.~Cuenda, ``{Vulnerability of overlay networks under
  malware spreading},'' \emph{arXiv preprint}, vol. abs/1310.0, 2013.

\bibitem{rsharmaDecInfoDissMulDim}
R.~Sharma and A.~Datta, ``{Decentralized Information Dissemination in
  Multidimensional Semantic Social Overlays},'' in \emph{Distributed Computing
  and Networking}, ser. Lecture Notes in Computer Science.\hskip 1em plus 0.5em
  minus 0.4em\relax Springer Berlin Heidelberg, 2012, vol. 7129, pp. 473--487.

\bibitem{IdnCritNodMulLayNwUnMulVecMalAtk}
J.~G. {Rafael Vida} and S.~Cuenda, ``{Identifying critical nodes in
  multi-layered networks under multi-vector malware attack},'' \emph{Int. J.
  Complex Systems in Science}, vol.~3, no.~1, pp. 97--105, 2013.

\bibitem{Wei2012}
\BIBentryALTinterwordspacing
X.~Wei, N.~Valler, B.~A. Prakash, I.~Neamtiu, M.~Faloutsos, and C.~Faloutsos,
  ``{Competing memes propagation on networks: a case study of composite
  networks},'' \emph{SIGCOMM Computer Communication Review}, vol.~42, no.~5,
  pp. 5--12, Oct. 2012. [Online]. Available:
  \url{http://dx.doi.org/10.1145/2378956.2378958}
\BIBentrySTDinterwordspacing

\bibitem{CompMemProOnNwPersp}
X.~Wei, N.~C. Valler, B.~A. Prakash, I.~Neamtiu, M.~Faloutsos, and
  C.~Faloutsos, ``{Competing Memes Propagation on Networks: A Network Science
  Perspective},'' \emph{IEEE Journal on Selected Areas in Communications},
  vol.~31, no.~6, pp. 1049--1060, Jun. 2013.

\bibitem{Budak2011}
\BIBentryALTinterwordspacing
C.~Budak, D.~Agrawal, and A.~{El Abbadi}, ``{Limiting the spread of
  misinformation in social networks},'' in \emph{International conference on
  World wide web}.\hskip 1em plus 0.5em minus 0.4em\relax New York, New York,
  USA: ACM Press, 2011, p. 665. [Online]. Available:
  \url{http://dblp.uni-trier.de/db/conf/www/www2011.html\#BudakAA11}
\BIBentrySTDinterwordspacing

\bibitem{Wang2011a}
\BIBentryALTinterwordspacing
Y.~Wang and G.~Xiao, ``{Effects of Interconnections on Epidemics in Network of
  Networks},'' in \emph{Conference on Wireless Communications, Networking and
  Mobile Computing}.\hskip 1em plus 0.5em minus 0.4em\relax Ieee, Sep. 2011,
  pp. 1--4. [Online]. Available:
  \url{http://ieeexplore.ieee.org/lpdocs/epic03/wrapper.htm?arnumber=6040146}
\BIBentrySTDinterwordspacing

\bibitem{Yagan2013}
\BIBentryALTinterwordspacing
O.~Yagan and D.~Qian, ``{Conjoining speeds up information diffusion in
  overlaying social-physical networks},'' \emph{IEEE Journal on Selected Areas
  in Communications (JSAC)}, vol.~31, no.~6, pp. 1038--1048, 2013. [Online].
  Available:
  \url{http://ieeexplore.ieee.org/xpls/abs\_all.jsp?arnumber=6517108}
\BIBentrySTDinterwordspacing

\bibitem{Saumell-Mendiola2012}
\BIBentryALTinterwordspacing
A.~Saumell-Mendiola, M.~A. Serrano, and M.~Bogu\~{n}\'{a}, ``{Epidemic
  spreading on interconnected networks},'' \emph{Physical Review E}, vol.~86,
  no.~2, p. 026106, Aug. 2012. [Online]. Available:
  \url{http://link.aps.org/doi/10.1103/PhysRevE.86.026106}
\BIBentrySTDinterwordspacing

\bibitem{Sanz2014}
J.~Sanz, C.-Y. Xia, S.~Meloni, and Y.~Moreno, ``{Dynamics of interacting
  diseases},'' Feb. 2014.

\bibitem{Hethcote2000}
H.~W. Hethcote, ``{The mathematics of infectious diseases},'' \emph{SIAM
  Review}, vol.~42, pp. 599--653, 2000.

\bibitem{Wang03}
\BIBentryALTinterwordspacing
Y.~Wang, D.~Chakrabarti, C.~Wang, and C.~Faloutsos, ``{Epidemic spreading in
  real networks: an eigenvalue viewpoint},'' in \emph{Reliable Distributed
  Systems, IEEE Symposium on}, vol.~0, Carnegie Mellon Univ., Pittsburgh, PA,
  USA.\hskip 1em plus 0.5em minus 0.4em\relax Los Alamitos, CA, USA: IEEE, Oct.
  2003, pp. 25--34. [Online]. Available:
  \url{http://www.cs.cmu.edu/~deepay/mywww/papers/srds03.pdf}
\BIBentrySTDinterwordspacing

\bibitem{Gleeson2011}
\BIBentryALTinterwordspacing
J.~P. Gleeson, ``{High-Accuracy Approximation of Binary-State Dynamics on
  Networks},'' \emph{Physical Review Letters}, vol. 107, no.~6, p. 068701, Aug.
  2011. [Online]. Available:
  \url{http://link.aps.org/doi/10.1103/PhysRevLett.107.068701}
\BIBentrySTDinterwordspacing

\bibitem{DarabiSahneh2014}
\BIBentryALTinterwordspacing
F.~{Darabi Sahneh} and C.~Scoglio, ``{Competitive epidemic spreading over
  arbitrary multilayer networks},'' \emph{Physical Review E}, vol.~89, no.~6,
  p. 062817, Jun. 2014. [Online]. Available:
  \url{http://link.aps.org/doi/10.1103/PhysRevE.89.062817}
\BIBentrySTDinterwordspacing

\bibitem{Gomez2013}
\BIBentryALTinterwordspacing
S.~G\'{o}mez, a.~D\'{\i}az-Guilera, J.~G\'{o}mez-Garde\~{n}es, C.~J.
  P\'{e}rez-Vicente, Y.~Moreno, and a.~Arenas, ``{Diffusion Dynamics on
  Multiplex Networks},'' \emph{Physical Review Letters}, vol. 110, no.~2, p.
  028701, Jan. 2013. [Online]. Available:
  \url{http://link.aps.org/doi/10.1103/PhysRevLett.110.028701}
\BIBentrySTDinterwordspacing

\bibitem{PhysRevE.79.036113}
\BIBentryALTinterwordspacing
A.~Allard, P.~P.-A. No\"{e}l, L.~L. Dub\'{e}, and B.~Pourbohloul,
  ``{Heterogeneous bond percolation on multitype networks with an application
  to epidemic dynamics},'' \emph{Physical Review E}, vol.~79, no.~3, p. 036113,
  Mar. 2009. [Online]. Available:
  \url{http://pre.aps.org/abstract/PRE/v79/i3/e036113
  http://link.aps.org/doi/10.1103/PhysRevE.79.036113}
\BIBentrySTDinterwordspacing

\bibitem{Datta2011GSG}
\BIBentryALTinterwordspacing
A.~Datta and R.~Sharma, ``{GoDisco: Selective Gossip Based Dissemination of
  Information in Social Community Based Overlays},'' in \emph{Proceedings of
  the 12th International Conference on Distributed Computing and Networking},
  ser. ICDCN'11.\hskip 1em plus 0.5em minus 0.4em\relax Berlin, Heidelberg:
  Springer-Verlag, 2011, pp. 227--238. [Online]. Available:
  \url{http://dl.acm.org/citation.cfm?id=1946143.1946163}
\BIBentrySTDinterwordspacing

\bibitem{Marceau2011}
\BIBentryALTinterwordspacing
V.~Marceau, P.~P.-A. No\"{e}l, L.~H\'{e}bert-Dufresne, A.~Allard, and L.~J.
  Dub\'{e}, ``{Modeling the dynamical interaction between epidemics on overlay
  networks},'' \emph{Physical Review E}, vol. 026105, p.~15, Mar. 2011.
  [Online]. Available: \url{http://arxiv.org/abs/1103.4059
  http://pre.aps.org/abstract/PRE/v84/i2/e026105}
\BIBentrySTDinterwordspacing

\bibitem{Kermack1927}
W.~O. Kermack and A.~G. McKendrick, ``{A contribution to the mathematical
  theory of epidemics.}'' \emph{Proceeding of Royal Society}, no. 115, 1927.

\bibitem{Newman2010}
M.~Newman, \emph{{Networks: An Introduction}}.\hskip 1em plus 0.5em minus
  0.4em\relax New York, NY, USA: Oxford University Press, Inc., 2010.

\bibitem{Goldenberg2001}
J.~Goldenberg, B.~Libai, and E.~Muller, ``{Talk of the Network: A Complex
  Systems Look at the Underlying Process of Word-of-Mouth},'' \emph{Marketing
  Letters}, vol.~12, no.~3, pp. 211--223, Aug. 2001.

\bibitem{PhysRevE.81.036118}
\BIBentryALTinterwordspacing
S.~Funk and V.~A.~A. Jansen, ``{Interacting epidemics on overlay networks.}''
  \emph{Physical review. E}, vol.~81, p. 036118, Mar. 2010. [Online].
  Available: \url{http://www.ncbi.nlm.nih.gov/pubmed/20365826}
\BIBentrySTDinterwordspacing

\bibitem{DarabiSahneh2013a}
\BIBentryALTinterwordspacing
F.~{Darabi Sahneh}, C.~Scoglio, and P.~{Van Mieghem}, ``{Generalized Epidemic
  Mean-Field Model for Spreading Processes Over Multilayer Complex Networks},''
  \emph{IEEE/ACM Transactions on Networking}, vol.~21, no.~5, pp. 1609--1620,
  Oct. 2013. [Online]. Available:
  \url{http://dl.acm.org/citation.cfm?id=2578911.2578931}
\BIBentrySTDinterwordspacing

\bibitem{Granovetter1978}
\BIBentryALTinterwordspacing
M.~Granovetter, ``{Threshold Models of Collective Behavior},'' \emph{The
  American journal of sociology}, vol.~83, no.~6, pp. 1420--1443, 1978.
  [Online]. Available:
  \url{http://www.stanfordlibrary.us/dept/soc/people/mgranovetter/documents/granthreshold.pdf}
\BIBentrySTDinterwordspacing

\bibitem{Watts2002}
D.~J. Watts, ``{A Simple Model of Global Cascades on Random Networks},'' in
  \emph{Proceedings of the National Academy of Sciences of the United States of
  America}, vol.~99, no.~9, Apr. 2002, pp. 5766--5771.

\bibitem{Centola2007a}
\BIBentryALTinterwordspacing
D.~Centola, V.~M. Egu\'{\i}luz, and M.~W. Macy, ``{Cascade dynamics of complex
  propagation},'' \emph{Physica A: Statistical Mechanics and its Applications},
  vol. 374, no.~1, pp. 449--456, Jan. 2007. [Online]. Available:
  \url{http://linkinghub.elsevier.com/retrieve/pii/S0378437106007679}
\BIBentrySTDinterwordspacing

\bibitem{Morris2000}
\BIBentryALTinterwordspacing
S.~Morris, ``{Contagion},'' \emph{The Review of Economic Studies}, vol.~67,
  no.~1, 2000. [Online]. Available:
  \url{http://papers.ssrn.com/abstract=234855}
\BIBentrySTDinterwordspacing

\bibitem{Brummitt2012a}
\BIBentryALTinterwordspacing
C.~D. Brummitt, K.-M. Lee, and K.-I. Goh, ``{Multiplexity-facilitated cascades
  in networks},'' \emph{Physical Review E}, vol.~85, no.~4, p. 045102, Apr.
  2012. [Online]. Available:
  \url{http://link.aps.org/doi/10.1103/PhysRevE.85.045102}
\BIBentrySTDinterwordspacing

\bibitem{Yagan2012}
\BIBentryALTinterwordspacing
O.~Yagan and V.~Gligor, ``{Analysis of complex contagions in random multiplex
  networks.}'' \emph{Physical review. E, Statistical, nonlinear, and soft
  matter physics}, vol.~86, no. 3 Pt 2, p. 036103, Sep. 2012. [Online].
  Available: \url{http://www.ncbi.nlm.nih.gov/pubmed/23030976}
\BIBentrySTDinterwordspacing

\bibitem{Kossinets2003}
G.~Kossinets, \emph{{Effects of missing data in social networks âˆ—}}, 2003.

\bibitem{Ramezanian2014}
\BIBentryALTinterwordspacing
R.~Ramezanian, M.~Salehi, M.~Magnani, and D.~Montesi, ``{Diffusion of
  Innovations over Multiplex Social Networks},'' Aug. 2014. [Online].
  Available: \url{http://arxiv-web3.library.cornell.edu/abs/1408.5806}
\BIBentrySTDinterwordspacing

\bibitem{Wilf2006}
\BIBentryALTinterwordspacing
H.~S. Wilf, \emph{{Generatingfunctionology}}.\hskip 1em plus 0.5em minus
  0.4em\relax A. K. Peters, Ltd., Jan. 2006. [Online]. Available:
  \url{http://dl.acm.org/citation.cfm?id=1204575}
\BIBentrySTDinterwordspacing

\bibitem{Resnick1992}
\BIBentryALTinterwordspacing
S.~I. Resnick, \emph{{Adventures in Stochastic Processes}}.\hskip 1em plus
  0.5em minus 0.4em\relax Birkhauser Verlag, 1992. [Online]. Available:
  \url{http://books.google.it/books/about/Adventures\_in\_Stochastic\_Processes.html?id=YGjTl8iX-HsC\&pgis=1}
\BIBentrySTDinterwordspacing

\bibitem{harris2002}
\BIBentryALTinterwordspacing
T.~E. Harris, \emph{{The Theory of Branching Processes}}.\hskip 1em plus 0.5em
  minus 0.4em\relax Dover Publications, 2002. [Online]. Available:
  \url{http://books.google.it/books/about/The\_Theory\_of\_Branching\_Processes.html?id=EbEliYYS3noC\&pgis=1}
\BIBentrySTDinterwordspacing

\bibitem{Kenah2007}
\BIBentryALTinterwordspacing
E.~Kenah and J.~Robins, ``{Second look at the spread of epidemics on
  networks},'' \emph{Physical Review E}, vol.~76, no.~3, p. 036113, Sep. 2007.
  [Online]. Available: \url{http://link.aps.org/doi/10.1103/PhysRevE.76.036113}
\BIBentrySTDinterwordspacing

\bibitem{Hebert-Dufresne2013}
\BIBentryALTinterwordspacing
L.~H\'{e}bert-Dufresne, O.~Patterson-Lomba, G.~M. Goerg, and B.~M. Althouse,
  ``{Pathogen mutation modeled by competition between site and bond
  percolation},'' \emph{Physical review letters}, vol. 110, no.~10, p. 108103,
  Mar. 2013. [Online]. Available:
  \url{http://www.ncbi.nlm.nih.gov/pubmed/23521302}
\BIBentrySTDinterwordspacing

\bibitem{Callaway2000}
\BIBentryALTinterwordspacing
D.~S. Callaway, M.~E.~J. Newman, S.~H. Strogatz, and D.~J. Watts, ``{Network
  Robustness and Fragility: Percolation on Random Graphs},'' \emph{Physical
  Review Letters}, vol.~85, no.~25, pp. 5468--5471, Dec. 2000. [Online].
  Available: \url{http://link.aps.org/doi/10.1103/PhysRevLett.85.5468}
\BIBentrySTDinterwordspacing

\bibitem{Baxter2014}
\BIBentryALTinterwordspacing
G.~J. Baxter, S.~N. Dorogovtsev, J.~F.~F. Mendes, and D.~Cellai, ``{Weak
  percolation on multiplex networks},'' \emph{Physical Review E}, vol.~89,
  no.~4, p. 042801, Apr. 2014. [Online]. Available:
  \url{http://link.aps.org/doi/10.1103/PhysRevE.89.042801}
\BIBentrySTDinterwordspacing

\bibitem{GomezSpreadingEPL2010}
\BIBentryALTinterwordspacing
S.~G\'{o}mez, A.~Arenas, J.~Borge-Holthoefer, S.~Meloni, and Y.~Moreno,
  ``{Discrete-time Markov chain approach to contact-based disease spreading in
  complex networks},'' \emph{EPL (Europhysics Letters)}, vol.~89, no.~3, Feb.
  2010. [Online]. Available: \url{http://dx.doi.org/10.1209/0295-5075/89/38009}
\BIBentrySTDinterwordspacing

\bibitem{Wu2014}
\BIBentryALTinterwordspacing
Q.~Wu, H.~Zhang, M.~Small, and X.~Fu, ``{Threshold analysis of the
  susceptible-infected-susceptible model on overlay networks},''
  \emph{Communications in Nonlinear Science and Numerical Simulation}, vol.~19,
  no.~7, pp. 2435--2443, Jul. 2014. [Online]. Available:
  \url{http://linkinghub.elsevier.com/retrieve/pii/S1007570413005674}
\BIBentrySTDinterwordspacing

\bibitem{Zinoviev2010}
\BIBentryALTinterwordspacing
D.~Zinoviev and V.~Duong, ``{A game theoretical approach to modeling
  full-duplex information dissemination},'' in \emph{Computer Simulation
  Conference}.\hskip 1em plus 0.5em minus 0.4em\relax Society for Computer
  Simulation International, Jul. 2010, pp. 358--363. [Online]. Available:
  \url{http://dl.acm.org/citation.cfm?id=1999416.1999462}
\BIBentrySTDinterwordspacing

\bibitem{Qiu2012}
\BIBentryALTinterwordspacing
W.~Qiu, Y.~Wang, and J.~Yu, ``{A game theoretical model of information
  dissemination in social network},'' in \emph{International Conference on
  Complex Systems (ICCS)}.\hskip 1em plus 0.5em minus 0.4em\relax IEEE, Nov.
  2012, pp. 1--6. [Online]. Available:
  \url{http://ieeexplore.ieee.org/lpdocs/epic03/wrapper.htm?arnumber=6458551}
\BIBentrySTDinterwordspacing

\bibitem{Banerjee2013}
\BIBentryALTinterwordspacing
A.~Banerjee, V.~Gauthier, H.~Labiod, and H.~Afifi, ``{Cooperation Optimized
  Design for Information Dissemination in Vehicular Networks using Evolutionary
  Game Theory},'' \emph{arXiv preprint}, Jan. 2013. [Online]. Available:
  \url{http://arxiv.org/abs/1301.1268}
\BIBentrySTDinterwordspacing

\bibitem{Sun2014635}
Y.~Sun, C.~Liu, C.-X. Zhang, and Z.-K. Zhang, ``{Epidemic spreading on weighted
  complex networks},'' \emph{Physics Letters A}, vol. 378, no.~78, pp.
  635--640, 2014.

\bibitem{Radicchi2013}
\BIBentryALTinterwordspacing
F.~Radicchi and A.~Arenas, ``{Abrupt transition in the structural formation of
  interconnected networks},'' \emph{Nature Physics}, vol.~9, no.~11, pp.
  717--720, Sep. 2013. [Online]. Available:
  \url{http://dx.doi.org/10.1038/nphys2761}
\BIBentrySTDinterwordspacing

\bibitem{Radicchi2014}
\BIBentryALTinterwordspacing
F.~Radicchi, ``{Driving Interconnected Networks to Supercriticality},''
  \emph{Physical Review X}, vol.~4, no.~2, p. 021014, Apr. 2014. [Online].
  Available: \url{http://link.aps.org/doi/10.1103/PhysRevX.4.021014}
\BIBentrySTDinterwordspacing

\bibitem{Dorogovtsev2003}
\BIBentryALTinterwordspacing
S.~N. Dorogovtsev and J.~F.~F. Mendes, ``{Evolution of Networks: From
  Biological Nets to the Internet and WWW (Physics)},'' Mar. 2003. [Online].
  Available: \url{http://dl.acm.org/citation.cfm?id=1212782}
\BIBentrySTDinterwordspacing

\bibitem{Sahneh2013b}
\BIBentryALTinterwordspacing
F.~Sahneh, C.~Scoglio, and F.~Chowdhury, ``{Effect of coupling on the epidemic
  threshold in interconnected complex networks: A spectral analysis},'' in
  \emph{American Control Conference (ACC)}, 2013, pp. 1--7. [Online].
  Available:
  \url{http://ieeexplore.ieee.org/xpls/abs\_all.jsp?arnumber=6580178}
\BIBentrySTDinterwordspacing

\bibitem{Parshani2010}
\BIBentryALTinterwordspacing
R.~Parshani, C.~Rozenblat, D.~Ietri, C.~Ducruet, and S.~Havlin,
  ``{Inter-similarity between coupled networks},'' \emph{EPL (Europhysics
  Letters)}, vol.~92, no.~6, p. 68002, Dec. 2010. [Online]. Available:
  \url{http://stacks.iop.org/0295-5075/92/i=6/a=68002}
\BIBentrySTDinterwordspacing

\bibitem{Qian2013}
D.~Qian, O.~Yagan, L.~Yang, and K.~Xing, ``{Diffusion of real-time information
  in overlaying social-physical networks: network coupling and clique
  structure},'' \emph{Network Science}, vol.~3, no. 1-4, pp. 43--53, 2013.

\bibitem{Soderberg2003a}
\BIBentryALTinterwordspacing
B.~S\"{o}derberg, ``{Properties of random graphs with hidden color},''
  \emph{Physical Review E}, vol.~68, no.~2, p. 026107, Aug. 2003. [Online].
  Available: \url{http://link.aps.org/doi/10.1103/PhysRevE.68.026107}
\BIBentrySTDinterwordspacing

\bibitem{DeDomenico2013a}
\BIBentryALTinterwordspacing
M.~{De Domenico}, A.~Sol\'{e}-Ribalta, E.~Omodei, S.~G\'{o}mez, and A.~Arenas,
  ``{Centrality in Interconnected Multilayer Networks},'' \emph{arXiv
  preprint}, p.~12, Nov. 2013. [Online]. Available:
  \url{http://arxiv.org/abs/1311.2906}
\BIBentrySTDinterwordspacing

\bibitem{Hu2014a}
\BIBentryALTinterwordspacing
Y.~Hu, S.~Havlin, and H.~A. Makse, ``{Conditions for Viral Influence Spreading
  through Multiplex Correlated Social Networks},'' \emph{Physical Review X},
  vol.~4, no.~2, p. 021031, May 2014. [Online]. Available:
  \url{http://link.aps.org/doi/10.1103/PhysRevX.4.021031}
\BIBentrySTDinterwordspacing

\bibitem{Shai2012}
\BIBentryALTinterwordspacing
S.~Shai and S.~Dobson, ``{Effect of resource constraints on intersimilar
  coupled networks},'' \emph{Physical Review E}, vol.~86, no.~6, p. 066120,
  Dec. 2012. [Online]. Available:
  \url{http://link.aps.org/doi/10.1103/PhysRevE.86.066120}
\BIBentrySTDinterwordspacing

\bibitem{Brockmann2006}
\BIBentryALTinterwordspacing
D.~Brockmann, L.~Hufnagel, and T.~Geisel, ``{The scaling laws of human
  travel},'' \emph{Nature}, vol. 439, no. 7075, pp. 462--5, Jan. 2006.
  [Online]. Available: \url{http://dx.doi.org/10.1038/nature04292}
\BIBentrySTDinterwordspacing

\bibitem{Miritello2011}
\BIBentryALTinterwordspacing
G.~Miritello, E.~Moro, and R.~Lara, ``{Dynamical strength of social ties in
  information spreading},'' \emph{Physical Review E}, vol.~83, no.~4, p.
  045102, Apr. 2011. [Online]. Available:
  \url{http://link.aps.org/doi/10.1103/PhysRevE.83.045102}
\BIBentrySTDinterwordspacing

\bibitem{Karsai2011}
\BIBentryALTinterwordspacing
M.~Karsai, M.~Kivel\"{a}, R.~K. Pan, K.~Kaski, J.~Kert\'{e}sz, A.-L.
  Barab\'{a}si, and J.~Saram\"{a}ki, ``{Small but slow world: How network
  topology and burstiness slow down spreading},'' \emph{Physical Review E},
  vol.~83, no.~2, p. 025102, Feb. 2011. [Online]. Available:
  \url{http://link.aps.org/doi/10.1103/PhysRevE.83.025102}
\BIBentrySTDinterwordspacing

\bibitem{Tang2011}
S.~Tang, J.~Yuan, X.~Mao, X.~Y. Li, W.~Chen, and G.~Dai, ``{Relationship
  classification in large scale online social networks and its impact on
  information propagation},'' in \emph{IEEE International Conference on
  Computer Communications (INFOCOM)}, 2011, pp. 2291--2299.

\bibitem{DeDomenico2014}
\BIBentryALTinterwordspacing
M.~{De Domenico}, A.~Sol\'{e}-Ribalta, S.~G\'{o}mez, and A.~Arenas,
  ``{Navigability of interconnected networks under random failures.}''
  \emph{Proceedings of the National Academy of Sciences of the United States of
  America}, vol. 111, no.~23, pp. 8351--6, Jun. 2014. [Online]. Available:
  \url{http://www.pnas.org/content/111/23/8351}
\BIBentrySTDinterwordspacing

\bibitem{Zhong2014}
C.~Zhong, M.~Salehi, S.~Shah, M.~Cobzarenco, N.~Sastry, and M.~Cha, ``{Social
  bootstrapping: how pinterest and last.fm social communities benefit by
  borrowing links from facebook},'' in \emph{World Wide Web}, Apr. 2014, pp.
  305--314.

\bibitem{Cellai2013}
\BIBentryALTinterwordspacing
D.~Cellai, E.~L\'{o}pez, J.~Zhou, J.~P. Gleeson, and G.~Bianconi,
  ``{Percolation in multiplex networks with overlap},'' \emph{Physical Review
  E}, vol.~88, no.~5, p. 052811, Nov. 2013. [Online]. Available:
  \url{http://link.aps.org/doi/10.1103/PhysRevE.88.052811}
\BIBentrySTDinterwordspacing

\bibitem{Szell2010}
\BIBentryALTinterwordspacing
M.~Szell, R.~Lambiotte, and S.~Thurner, ``{Multirelational organization of
  large-scale social networks in an online world.}'' \emph{Proceedings of the
  National Academy of Sciences of the United States of America}, vol. 107,
  no.~31, pp. 13\,636--41, Aug. 2010. [Online]. Available:
  \url{http://www.pnas.org/content/107/31/13636.full}
\BIBentrySTDinterwordspacing

\bibitem{Bianconi2013}
\BIBentryALTinterwordspacing
G.~Bianconi, ``{Statistical mechanics of multiplex networks: Entropy and
  overlap},'' \emph{Physical Review E}, vol.~87, no.~6, p. 062806, Jun. 2013.
  [Online]. Available: \url{http://link.aps.org/doi/10.1103/PhysRevE.87.062806}
\BIBentrySTDinterwordspacing

\bibitem{Karrer2011}
\BIBentryALTinterwordspacing
B.~Karrer and M.~Newman, ``{Competing epidemics on complex networks},''
  \emph{Physical Review E}, vol.~84, no.~3, p. 036106, Sep. 2011. [Online].
  Available: \url{http://link.aps.org/doi/10.1103/PhysRevE.84.036106}
\BIBentrySTDinterwordspacing

\bibitem{Prakash2012}
\BIBentryALTinterwordspacing
B.~A. Prakash, A.~Beutel, R.~Rosenfeld, and C.~Faloutsos, ``{Winner takes all:
  competing viruses or ideas on fair-play networks},'' in \emph{International
  conference on World Wide Web - WWW '12}.\hskip 1em plus 0.5em minus
  0.4em\relax New York, New York, USA: ACM Press, Apr. 2012, p. 1037. [Online].
  Available: \url{http://dl.acm.org/citation.cfm?id=2187836.2187975}
\BIBentrySTDinterwordspacing

\bibitem{VandeBovenkamp2014}
\BIBentryALTinterwordspacing
R.~van~de Bovenkamp, F.~Kuipers, and P.~{Van Mieghem}, ``{Domination-time
  dynamics in susceptible-infected-susceptible virus competition on
  networks},'' \emph{Physical Review E}, vol.~89, no.~4, p. 042818, Apr. 2014.
  [Online]. Available: \url{http://link.aps.org/doi/10.1103/PhysRevE.89.042818}
\BIBentrySTDinterwordspacing

\bibitem{Kostka2008}
\BIBentryALTinterwordspacing
J.~Kostka, Y.~A. Oswald, and R.~Wattenhofer, \emph{{Structural Information and
  Communication Complexity}}, ser. Lecture Notes in Computer Science, A.~A.
  Shvartsman and P.~Felber, Eds.\hskip 1em plus 0.5em minus 0.4em\relax Berlin,
  Heidelberg: Springer Berlin Heidelberg, Jun. 2008, vol. 5058. [Online].
  Available: \url{http://dl.acm.org/citation.cfm?id=1424630.1424648}
\BIBentrySTDinterwordspacing

\bibitem{Goyal2012}
\BIBentryALTinterwordspacing
S.~Goyal and M.~Kearns, ``{Competitive contagion in networks},'' in
  \emph{Symposium on Theory of Computing - STOC '12}.\hskip 1em plus 0.5em
  minus 0.4em\relax New York, New York, USA: ACM Press, May 2012, p. 759.
  [Online]. Available: \url{http://dl.acm.org/citation.cfm?id=2213977.2214046}
\BIBentrySTDinterwordspacing

\bibitem{Tzoumas2012}
\BIBentryALTinterwordspacing
V.~Tzoumas, C.~Amanatidis, and E.~Markakis, ``{Internet and Network
  Economics},'' in \emph{International conference on Internet and Network
  Economics}, ser. Lecture Notes in Computer Science, P.~W. Goldberg, Ed., vol.
  7695.\hskip 1em plus 0.5em minus 0.4em\relax Berlin, Heidelberg: Springer
  Berlin Heidelberg, Dec. 2012, pp. 1--14. [Online]. Available:
  \url{http://dl.acm.org/citation.cfm?id=2436756.2436758}
\BIBentrySTDinterwordspacing

\bibitem{Fazeli2012}
\BIBentryALTinterwordspacing
A.~Fazeli and A.~Jadbabaie, ``\BIBforeignlanguage{English}{{Game theoretic
  analysis of a strategic model of competitive contagion and product adoption
  in social networks}},'' in \emph{\BIBforeignlanguage{English}{IEEE Conference
  on Decision and Control (CDC)}}.\hskip 1em plus 0.5em minus 0.4em\relax IEEE,
  Dec. 2012, pp. 74--79. [Online]. Available:
  \url{http://ieeexplore.ieee.org/articleDetails.jsp?arnumber=6426222}
\BIBentrySTDinterwordspacing

\bibitem{Bharathi2007}
\BIBentryALTinterwordspacing
S.~Bharathi, D.~Kempe, and M.~Salek, ``{Competitive influence maximization in
  social networks},'' in \emph{International conference on Internet and network
  economics}.\hskip 1em plus 0.5em minus 0.4em\relax Springer-Verlag, Dec.
  2007, pp. 306--311. [Online]. Available:
  \url{http://dl.acm.org/citation.cfm?id=1781894.1781932}
\BIBentrySTDinterwordspacing

\bibitem{Dubey2006}
\BIBentryALTinterwordspacing
P.~Dubey, R.~Garg, and B.~{De Meyer}, ``{Internet and Network Economics},'' in
  \emph{International conference on Internet and Network Economics}, ser.
  Lecture Notes in Computer Science, P.~Spirakis, M.~Mavronicolas, and
  S.~Kontogiannis, Eds., vol. 4286.\hskip 1em plus 0.5em minus 0.4em\relax
  Berlin, Heidelberg: Springer Berlin Heidelberg, Dec. 2006, pp. 162--173.
  [Online]. Available: \url{http://dl.acm.org/citation.cfm?id=2081411.2081427}
\BIBentrySTDinterwordspacing

\bibitem{Newman2005}
\BIBentryALTinterwordspacing
M.~Newman, ``{Threshold Effects for Two Pathogens Spreading on a Network},''
  \emph{Physical Review Letters}, vol.~95, no.~10, Sep. 2005. [Online].
  Available: \url{http://dx.doi.org/10.1103/physrevlett.95.108701}
\BIBentrySTDinterwordspacing

\bibitem{Rogers2003}
\BIBentryALTinterwordspacing
E.~M. Rogers, \emph{{Diffusion of innovations}}, 5th~ed.\hskip 1em plus 0.5em
  minus 0.4em\relax Free Press, 2003. [Online]. Available:
  \url{http://books.google.it/books/about/Diffusion\_of\_innovations.html?id=zw0-AAAAIAAJ\&pgis=1}
\BIBentrySTDinterwordspacing

\bibitem{Immorlica2007}
\BIBentryALTinterwordspacing
N.~Immorlica, J.~Kleinberg, M.~Mahdian, and T.~Wexler, ``{The role of
  compatibility in the diffusion of technologies through social networks},'' in
  \emph{ACM conference on Electronic commerce - EC '07}.\hskip 1em plus 0.5em
  minus 0.4em\relax New York, New York, USA: ACM Press, Jun. 2007, p.~75.
  [Online]. Available: \url{http://dl.acm.org/citation.cfm?id=1250910.1250923}
\BIBentrySTDinterwordspacing

\bibitem{MatthewO.Jackson2005}
M.~O. Jackson and Y.~Leeat, ``{Diffusion on Social Networks},'' \emph{Economie
  Publique}, vol.~16, no.~2, pp. 69--82, 2005.

\bibitem{Lee2012}
K.-M. Lee, J.~Y. Kim, W.-k. Cho, K.-I. Goh, and I.-M. Kim, ``{Correlated
  multiplexity and connectivity of multiplex random networks},'' \emph{New
  Journal of Physics}, vol.~14, p. 033027, 2012.

\bibitem{SeedSizeStrAffectCascad}
C.~D. {Gleeson J}, ``{Seed size strongly affects cascades on random
  networks},'' in \emph{Phys Rev E Stat Nonlin Soft Matter Phys}, vol.~75, no.
  5 Pt 2.\hskip 1em plus 0.5em minus 0.4em\relax Physical Review E, 2007, p.
  056103.

\bibitem{Domingos:2001:MNV:502512.502525}
\BIBentryALTinterwordspacing
P.~Domingos and M.~Richardson, ``{Mining the Network Value of Customers},'' in
  \emph{ACM SIGKDD International Conference on Knowledge Discovery and Data
  Mining}, ser. KDD '01.\hskip 1em plus 0.5em minus 0.4em\relax New York, NY,
  USA: ACM, 2001, pp. 57--66. [Online]. Available:
  \url{http://doi.acm.org/10.1145/502512.502525}
\BIBentrySTDinterwordspacing

\bibitem{Leskovec2007}
\BIBentryALTinterwordspacing
J.~Leskovec, A.~Krause, C.~Guestrin, C.~Faloutsos, J.~VanBriesen, and
  N.~Glance, ``{Cost-effective outbreak detection in networks},''
  \emph{Proceedings of the 13th ACM SIGKDD international conference on
  Knowledge discovery and data mining - KDD '07}, p. 420, 2007. [Online].
  Available: \url{http://portal.acm.org/citation.cfm?doid=1281192.1281239}
\BIBentrySTDinterwordspacing

\bibitem{Kempe2005}
\BIBentryALTinterwordspacing
D.~Kempe, J.~Kleinberg, and {\'E}.~Tardos,
  ``\BIBforeignlanguage{English}{Influential nodes in a diffusion model for
  social networks},'' in \emph{\BIBforeignlanguage{English}{Automata, Languages
  and Programming}}, ser. Lecture Notes in Computer Science, L.~Caires, G.~F.
  Italiano, L.~Monteiro, C.~Palamidessi, and M.~Yung, Eds.\hskip 1em plus 0.5em
  minus 0.4em\relax Springer Berlin Heidelberg, 2005, vol. 3580, pp.
  1127--1138. [Online]. Available: \url{http://dx.doi.org/10.1007/11523468_91}
\BIBentrySTDinterwordspacing

\bibitem{Chen:2009:EIM:1557019.1557047}
\BIBentryALTinterwordspacing
W.~Chen, Y.~Wang, and S.~Yang, ``{Efficient Influence Maximization in Social
  Networks},'' in \emph{Proceedings of the 15th ACM SIGKDD International
  Conference on Knowledge Discovery and Data Mining}, ser. KDD '09.\hskip 1em
  plus 0.5em minus 0.4em\relax New York, NY, USA: ACM, 2009, pp. 199--208.
  [Online]. Available: \url{http://doi.acm.org/10.1145/1557019.1557047}
\BIBentrySTDinterwordspacing

\bibitem{Chen:2010:SIM:1835804.1835934}
\BIBentryALTinterwordspacing
W.~Chen, C.~Wang, and Y.~Wang, ``{Scalable Influence Maximization for Prevalent
  Viral Marketing in Large-scale Social Networks},'' in \emph{Proceedings of
  the 16th ACM SIGKDD International Conference on Knowledge Discovery and Data
  Mining}, ser. KDD '10.\hskip 1em plus 0.5em minus 0.4em\relax New York, NY,
  USA: ACM, 2010, pp. 1029--1038. [Online]. Available:
  \url{http://doi.acm.org/10.1145/1835804.1835934}
\BIBentrySTDinterwordspacing

\bibitem{DiscInfAuthHetAcdNwCoMet}
\BIBentryALTinterwordspacing
Q.~Meng and P.~J. Kennedy, ``{Discovering Influential Authors in Heterogeneous
  Academic Networks by a Co-ranking Method},'' in \emph{ACM International
  Conference on Conference on Information and Knowledge Management (CIKM)},
  ser. CIKM '13.\hskip 1em plus 0.5em minus 0.4em\relax New York, NY, USA: ACM,
  2013, pp. 1029--1036. [Online]. Available:
  \url{http://doi.acm.org/10.1145/2505515.2505534}
\BIBentrySTDinterwordspacing

\bibitem{Li2012a}
\BIBentryALTinterwordspacing
C.~Li, J.~Luo, J.~J. Huang, and J.~Fan, ``{Multi-Layer Network for Influence
  Propagation over Microblog},'' in \emph{Intelligence and Security
  Informatics}, ser. Lecture Notes in Computer Science, M.~Chau, G.~Wang,
  W.~Yue, and H.~Chen, Eds.\hskip 1em plus 0.5em minus 0.4em\relax Springer
  Berlin Heidelberg, 2012, vol. 7299, pp. 60--72. [Online]. Available:
  \url{http://link.springer.com/chapter/10.1007/978-3-642-30428-6\_5}
\BIBentrySTDinterwordspacing

\bibitem{Magnani2013a}
\BIBentryALTinterwordspacing
M.~Magnani, B.~Micenkov\'{a}, and L.~R. 0003, ``{Combinatorial Analysis of
  Multiple Networks},'' \emph{arXiv preprint}, vol. abs/1303.4, 2013. [Online].
  Available:
  \url{http://dblp.uni-trier.de/db/journals/corr/corr1303.html\#abs-1303-4986}
\BIBentrySTDinterwordspacing

\bibitem{Sole-Ribalta2014}
\BIBentryALTinterwordspacing
A.~Sol\'{e}-Ribalta, M.~{De Domenico}, S.~G\'{o}mez, and A.~Arenas,
  ``{Centrality rankings in multiplex networks},'' in \emph{ACM conference on
  Web science - WebSci}.\hskip 1em plus 0.5em minus 0.4em\relax New York, New
  York, USA: ACM Press, Jun. 2014, pp. 149--155. [Online]. Available:
  \url{http://dl.acm.org/citation.cfm?id=2615569.2615687}
\BIBentrySTDinterwordspacing

\bibitem{Sola2013}
\BIBentryALTinterwordspacing
L.~Sol\'{a}, M.~Romance, R.~Criado, J.~Flores, A.~{Garc\'{\i}a del Amo}, and
  S.~Boccaletti, ``{Multiplex PageRank},'' \emph{Chaos (Woodbury, N.Y.)},
  vol.~23, no.~3, p. 033131, Sep. 2013. [Online]. Available:
  \url{http://www.ncbi.nlm.nih.gov/pubmed/24089967}
\BIBentrySTDinterwordspacing

\bibitem{Halu2013}
\BIBentryALTinterwordspacing
A.~Halu, R.~Mondrag\'{o}n, P.~Panzarasa, and G.~Bianconi, ``{Multiplex
  PageRank},'' \emph{PloS one}, pp. 1--16, 2013. [Online]. Available:
  \url{http://dx.plos.org/10.1371/journal.pone.0078293.g006}
\BIBentrySTDinterwordspacing

\bibitem{Zhao2014}
\BIBentryALTinterwordspacing
D.~Zhao, L.~Li, S.~Li, Y.~Huo, and Y.~Yang, ``{Identifying influential
  spreaders in interconnected networks},'' \emph{Physica Scripta}, vol.~89,
  no.~1, p. 015203, Jan. 2014. [Online]. Available:
  \url{http://stacks.iop.org/1402-4896/89/i=1/a=015203?key=crossref.891f7f5fe61aab972aaa35ae7b58b002}
\BIBentrySTDinterwordspacing

\bibitem{Kitsak2010}
\BIBentryALTinterwordspacing
M.~Kitsak, L.~K. Gallos, S.~Havlin, F.~Liljeros, L.~Muchnik, H.~E. Stanley, and
  H.~A. Makse, ``{Identification of influential spreaders in complex
  networks},'' \emph{Nature Physics}, vol.~6, no.~11, pp. 888--893, Aug. 2010.
  [Online]. Available: \url{http://dx.doi.org/10.1038/nphys1746}
\BIBentrySTDinterwordspacing

\bibitem{Jo2006}
\BIBentryALTinterwordspacing
H.-H. Jo, S.~{Ki Baek}, and H.-T. Moon, ``{Immunization dynamics on a two-layer
  network model},'' \emph{Physica A: Statistical Mechanics and its
  Applications}, vol. 361, no.~2, pp. 534--542, Mar. 2006. [Online]. Available:
  \url{http://dx.doi.org/10.1016/j.physa.2005.06.074}
\BIBentrySTDinterwordspacing

\bibitem{Funk2010a}
\BIBentryALTinterwordspacing
S.~Funk, E.~Gilad, and V.~a.~a. Jansen, ``{Endemic disease, awareness, and
  local behavioural response.}'' \emph{Journal of theoretical biology}, vol.
  264, no.~2, pp. 501--9, May 2010. [Online]. Available:
  \url{http://www.ncbi.nlm.nih.gov/pubmed/20184901}
\BIBentrySTDinterwordspacing

\bibitem{Funk2009}
\BIBentryALTinterwordspacing
S.~Funk, E.~Gilad, C.~Watkins, and V.~a.~a. Jansen, ``{The spread of awareness
  and its impact on epidemic outbreaks.}'' \emph{Proceedings of the National
  Academy of Sciences of the United States of America}, vol. 106, no.~16, pp.
  6872--7, Apr. 2009. [Online]. Available:
  \url{http://www.pubmedcentral.nih.gov/articlerender.fcgi?artid=2672559\&tool=pmcentrez\&rendertype=abstract}
\BIBentrySTDinterwordspacing

\bibitem{ModImpUsrAwrImmStrMosi}
B.~Mirzasoleiman, H.~R. Rabiee, and M.~Salehi, ``{Modeling the Impact of User
  Awareness on Immunization Strategies},'' in \emph{NetSciCom}, 2014.

\bibitem{Shai2013}
\BIBentryALTinterwordspacing
S.~Shai and S.~Dobson, ``{Coupled adaptive complex networks},'' \emph{Physical
  Review E}, vol.~87, no.~4, p. 042812, Apr. 2013. [Online]. Available:
  \url{http://link.aps.org/doi/10.1103/PhysRevE.87.042812}
\BIBentrySTDinterwordspacing

\bibitem{LDPM13:exploiting}
{Antonio Lima}, {Manlio De Domenico}, {Veljko Pejovic}, and {Mirco Musolesi},
  ``{Exploiting Cellular Data for Disease Containment and Information Campaigns
  Strategies in Country-wide Epidemics},'' in \emph{International Conference on
  the Analysis of Mobile Phone Datasets (NetMob'13)}, Boston, 2013.

\bibitem{Granell2013a}
\BIBentryALTinterwordspacing
C.~Granell, S.~G\'{o}mez, and A.~Arenas, ``{Dynamical Interplay between
  Awareness and Epidemic Spreading in Multiplex Networks},'' \emph{Physical
  Review Letters}, vol. 111, no.~12, p. 128701, Sep. 2013. [Online]. Available:
  \url{http://link.aps.org/doi/10.1103/PhysRevLett.111.128701}
\BIBentrySTDinterwordspacing

\bibitem{Granell2014}
\BIBentryALTinterwordspacing
------, ``{Competing spreading processes on multiplex networks: Awareness and
  epidemics},'' \emph{Physical Review E}, vol.~90, no.~1, p. 012808, Jul. 2014.
  [Online]. Available: \url{http://link.aps.org/doi/10.1103/PhysRevE.90.012808}
\BIBentrySTDinterwordspacing

\bibitem{Bagnoli2014}
\BIBentryALTinterwordspacing
F.~Bagnoli and E.~Massaro, ``\BIBforeignlanguage{English}{Risk perception and
  epidemic spreading in multiplex networks},'' in
  \emph{\BIBforeignlanguage{English}{ISCS 2014: Interdisciplinary Symposium on
  Complex Systems}}, ser. Emergence, Complexity and Computation, A.~Sanayei,
  O.~E.~RÃ¶ssler, and I.~Zelinka, Eds.\hskip 1em plus 0.5em minus 0.4em\relax
  Springer International Publishing, 2015, vol.~14, pp. 319--332. [Online].
  Available: \url{http://dx.doi.org/10.1007/978-3-319-10759-2_33}
\BIBentrySTDinterwordspacing

\bibitem{Funk2010c}
\BIBentryALTinterwordspacing
S.~Funk, M.~Salath\'{e}, and V.~A. Jansen, ``{Modelling the influence of human
  behaviour on the spread of infectious diseases: a review.}'' \emph{Journal of
  the Royal Society, Interface / the Royal Society}, vol.~7, no.~50, pp.
  1247--1256, Sep. 2010. [Online]. Available:
  \url{http://dx.doi.org/10.1098/rsif.2010.0142}
\BIBentrySTDinterwordspacing

\bibitem{Sahneh2012}
\BIBentryALTinterwordspacing
F.~Sahneh and C.~Scoglio, ``{Optimal information dissemination in epidemic
  networks},'' \emph{\ldots and Control (CDC), 2012 IEEE 51st \ldots}, pp.
  1657--1662, 2012. [Online]. Available:
  \url{http://ieeexplore.ieee.org/xpls/abs\_all.jsp?arnumber=6425833}
\BIBentrySTDinterwordspacing

\bibitem{Sahneh2014}
\BIBentryALTinterwordspacing
F.~Sahneh, F.~Chowdhury, G.~Brase, and C.~Scoglio,
  ``\BIBforeignlanguage{English}{{Individual-based Information Dissemination in
  Multilayer Epidemic Modeling}},''
  \emph{\BIBforeignlanguage{English}{Mathematical Modelling of Natural
  Phenomena}}, vol.~9, no.~2, pp. 136--152, Apr. 2014. [Online]. Available:
  \url{http://www.mmnp-journal.org/10.1051/mmnp/20149209}
\BIBentrySTDinterwordspacing

\bibitem{DarabiSahneh2011}
\BIBentryALTinterwordspacing
F.~{Darabi Sahneh} and C.~Scoglio, ``{Epidemic spread in human networks},'' in
  \emph{IEEE Conference on Decision and Control and European Control
  Conference}.\hskip 1em plus 0.5em minus 0.4em\relax IEEE, Dec. 2011, pp.
  3008--3013. [Online]. Available:
  \url{http://ieeexplore.ieee.org/lpdocs/epic03/wrapper.htm?arnumber=6161529}
\BIBentrySTDinterwordspacing

\bibitem{Sahneh2012a}
\BIBentryALTinterwordspacing
F.~D. Sahneh, F.~N. Chowdhury, and C.~M. Scoglio, ``{On the existence of a
  threshold for preventive behavioral responses to suppress epidemic
  spreading.}'' \emph{Scientific reports}, vol.~2, p. 632, Jan. 2012. [Online].
  Available:
  \url{http://www.pubmedcentral.nih.gov/articlerender.fcgi?artid=3433689\&tool=pmcentrez\&rendertype=abstract}
\BIBentrySTDinterwordspacing

\bibitem{dtn}
\BIBentryALTinterwordspacing
K.~Fall, ``A delay-tolerant network architecture for challenged internets,'' in
  \emph{Proceedings of the 2003 Conference on Applications, Technologies,
  Architectures, and Protocols for Computer Communications}, ser. SIGCOMM
  '03.\hskip 1em plus 0.5em minus 0.4em\relax New York, NY, USA: ACM, 2003, pp.
  27--34. [Online]. Available: \url{http://doi.acm.org/10.1145/863955.863960}
\BIBentrySTDinterwordspacing

\bibitem{dtn-routing}
\BIBentryALTinterwordspacing
S.~Jain, K.~Fall, and R.~Patra, ``Routing in a delay tolerant network,''
  \emph{SIGCOMM Comput. Commun. Rev.}, vol.~34, no.~4, pp. 145--158, Aug. 2004.
  [Online]. Available: \url{http://doi.acm.org/10.1145/1030194.1015484}
\BIBentrySTDinterwordspacing

\bibitem{Stopczynski2014}
\BIBentryALTinterwordspacing
A.~Stopczynski, V.~Sekara, P.~Sapiezynski, A.~Cuttone, M.~M. Madsen, J.~E.
  Larsen, and S.~Lehmann, ``{Measuring large-scale social networks with high
  resolution.}'' \emph{PloS one}, vol.~9, no.~4, p. e95978, Jan. 2014.
  [Online]. Available: \url{http://dx.plos.org/10.1371/journal.pone.0095978}
\BIBentrySTDinterwordspacing

\bibitem{Costa2005}
L.~D.~F. Costa, F.~A. Rodrigues, G.~Travieso, and P.~R.~V. Boas,
  ``{Characterization of complex networks: A survey of measurements},''
  \emph{Advances in Physics}, vol.~56, pp. 167--242, 2005.

\bibitem{Battiston2014}
\BIBentryALTinterwordspacing
F.~Battiston, V.~Nicosia, and V.~Latora, ``{Structural measures for multiplex
  networks},'' \emph{Physical Review E}, vol.~89, no.~3, p. 032804, Mar. 2014.
  [Online]. Available: \url{http://link.aps.org/doi/10.1103/PhysRevE.89.032804}
\BIBentrySTDinterwordspacing

\bibitem{Sun2012}
Y.~Sun and J.~Han, \emph{{Mining Heterogeneous Information Networks: Principles
  and Methodologies}}, ser. Synthesis Lectures on Data Mining and Knowledge
  Discovery.\hskip 1em plus 0.5em minus 0.4em\relax Morgan \{\&\} Claypool
  Publishers, 2012.

\bibitem{Beck2014}
F.~Beck, M.~Burch, S.~Diehl, and D.~Weiskopf, ``{The State of the Art in
  Visualizing Dynamic Graphs},'' in \emph{EuroVis STAR}, 2014.

\bibitem{de2014multilayer}
M.~{De Domenico}, M.~A. Porter, and A.~Arenas, ``{Multilayer Analysis and
  Visualization of Networks},'' \emph{arXiv preprint}, 2014.

\bibitem{Holme2012}
\BIBentryALTinterwordspacing
P.~Holme and J.~Saram\"{a}ki, ``{Temporal networks},'' \emph{Physics Reports},
  vol. 519, no.~3, pp. 97--125, Oct. 2012. [Online]. Available:
  \url{http://www.sciencedirect.com/science/article/pii/S0370157312000841}
\BIBentrySTDinterwordspacing

\bibitem{Vazquez2007}
\BIBentryALTinterwordspacing
A.~Vazquez, B.~R\'{a}cz, A.~Luk\'{a}cs, and A.-L. Barab\'{a}si, ``{Impact of
  Non-Poissonian Activity Patterns on Spreading Processes},'' \emph{Physical
  Review Letters}, vol.~98, no.~15, p. 158702, Apr. 2007. [Online]. Available:
  \url{http://link.aps.org/doi/10.1103/PhysRevLett.98.158702}
\BIBentrySTDinterwordspacing

\bibitem{Volz2009}
\BIBentryALTinterwordspacing
E.~Volz and L.~A. Meyers, ``{Epidemic thresholds in dynamic contact
  networks.}'' \emph{Journal of the Royal Society, Interface / the Royal
  Society}, vol.~6, no.~32, pp. 233--41, Mar. 2009. [Online]. Available:
  \url{http://rsif.royalsocietypublishing.org/content/6/32/233.abstract}
\BIBentrySTDinterwordspacing

\bibitem{Taylor2012}
\BIBentryALTinterwordspacing
M.~Taylor, T.~J. Taylor, and I.~Z. Kiss, ``{Epidemic threshold and control in a
  dynamic network},'' \emph{Physical Review E}, vol.~85, no.~1, p. 016103, Jan.
  2012. [Online]. Available:
  \url{http://link.aps.org/doi/10.1103/PhysRevE.85.016103}
\BIBentrySTDinterwordspacing

\bibitem{Gauvin2013}
\BIBentryALTinterwordspacing
L.~Gauvin, A.~Panisson, C.~Cattuto, and A.~Barrat,
  ``\BIBforeignlanguage{en}{{Activity clocks: spreading dynamics on temporal
  networks of human contact.}}'' \emph{\BIBforeignlanguage{en}{Scientific
  reports}}, vol.~3, p. 3099, Jan. 2013. [Online]. Available:
  \url{http://www.nature.com/srep/2013/131031/srep03099/full/srep03099.html}
\BIBentrySTDinterwordspacing

\bibitem{DBLP:journals/socnet/SnijdersLT13}
T.~A.~B. Snijders, A.~Lomi, and V.~J. Torl\'{o}, ``{A model for the multiplex
  dynamics of two-mode and one-mode networks, with an application to employment
  preference, friendship, and advice},'' \emph{Social Networks}, vol.~35,
  no.~2, pp. 265--276, 2013.

\bibitem{Valdano2014}
\BIBentryALTinterwordspacing
E.~Valdano, L.~Ferreri, C.~Poletto, and V.~Colizza, ``{Analytical computation
  of the epidemic threshold on temporal networks},'' \emph{arXiv preprint},
  p.~19, Jun. 2014. [Online]. Available: \url{http://arxiv.org/abs/1406.4815}
\BIBentrySTDinterwordspacing

\bibitem{Gross2006}
\BIBentryALTinterwordspacing
T.~Gross, C.~Dâ€™Lima, and B.~Blasius, ``{Epidemic Dynamics on an Adaptive
  Network},'' \emph{Physical Review Letters}, vol.~96, no.~20, p. 208701, May
  2006. [Online]. Available:
  \url{http://link.aps.org/doi/10.1103/PhysRevLett.96.208701}
\BIBentrySTDinterwordspacing

\bibitem{Gross2008}
\BIBentryALTinterwordspacing
T.~Gross and B.~Blasius, ``{Adaptive coevolutionary networks: a review},''
  \emph{Journal of the Royal Society, Interface / the Royal Society}, vol.~5,
  no.~20, pp. 259--71, Mar. 2008. [Online]. Available:
  \url{http://rsif.royalsocietypublishing.org/content/5/20/259.full}
\BIBentrySTDinterwordspacing

\bibitem{Meloni2011}
\BIBentryALTinterwordspacing
S.~Meloni, N.~Perra, A.~Arenas, S.~G\'{o}mez, Y.~Moreno, and A.~Vespignani,
  ``\BIBforeignlanguage{en}{{Modeling human mobility responses to the
  large-scale spreading of infectious diseases.}}''
  \emph{\BIBforeignlanguage{en}{Scientific reports}}, vol.~1, p.~62, Jan. 2011.
  [Online]. Available:
  \url{http://www.nature.com/srep/2011/110812/srep00062/full/srep00062.html}
\BIBentrySTDinterwordspacing

\end{thebibliography}
